\pdfoutput=1
\documentclass[12pt,a4paper]{article}

\usepackage{ifthen} 
\newboolean{pdflatex}
\setboolean{pdflatex}{true} 

\newboolean{articletitles}
\setboolean{articletitles}{true} 

\newboolean{uprightparticles}
\setboolean{uprightparticles}{false} 

\def\paperauthors{LHCb collaboration}
\def\paperasciititle{Template for writing LHCb papers} 
\def\papertitle{Measurement of the CKM angle $\gamma$ in the \BDKst channel using self-conjugate \Dhh decays} 
\def\paperkeywords{{High Energy Physics}, {LHCb}} 
\def\papercopyright{\the\year\ CERN for the benefit of the LHCb collaboration} 
\def\paperlicence{CC BY 4.0 licence}
\def\paperlicenceurl{https://creativecommons.org/licenses/by/4.0/}

\usepackage[top=1in, bottom=1.25in, left=1in, right=1in]{geometry}
\usepackage{booktabs} 
\usepackage{multirow}
\usepackage{adjustbox}
\usepackage{mathtools}
\usepackage{microtype}
\usepackage{lineno}  
\usepackage{xspace} 
\usepackage{caption} 

\usepackage{graphicx}  
\usepackage{color}
\usepackage{colortbl}
\graphicspath{{./figs/}} 

\usepackage{amsmath} 
\usepackage{amssymb}
\usepackage{amsfonts}
\usepackage{upgreek} 

\newcommand*\patchAmsMathEnvironmentForLineno[1]{%
\expandafter\let\csname old#1\expandafter\endcsname\csname #1\endcsname
\expandafter\let\csname oldend#1\expandafter\endcsname\csname
end#1\endcsname
 \renewenvironment{#1}%
   {\linenomath\csname old#1\endcsname}%
   {\csname oldend#1\endcsname\endlinenomath}%
}
\newcommand*\patchBothAmsMathEnvironmentsForLineno[1]{%
  \patchAmsMathEnvironmentForLineno{#1}%
  \patchAmsMathEnvironmentForLineno{#1*}%
}
\AtBeginDocument{%
\patchBothAmsMathEnvironmentsForLineno{equation}%
\patchBothAmsMathEnvironmentsForLineno{align}%
\patchBothAmsMathEnvironmentsForLineno{flalign}%
\patchBothAmsMathEnvironmentsForLineno{alignat}%
\patchBothAmsMathEnvironmentsForLineno{gather}%
\patchBothAmsMathEnvironmentsForLineno{multline}%
\patchBothAmsMathEnvironmentsForLineno{eqnarray}%
}

\def\rB {\ensuremath{r_\Bz}\xspace}
\def\deltaB {\ensuremath{\delta_\Bz}\xspace}
\def\BDKst {\ensuremath{ {\Bz\to\D\Kstarz}}\xspace}
\def\BsDKst {\ensuremath{ {\Bsb\to\Dz\Kstarz}}\xspace}
\def\BDh {\ensuremath{B^\pm\to\D h^\pm}\xspace}

\def\BKshhKpi {\ensuremath{\Bz\to(\KS h^+ h^-)_D\Kp\pim}\xspace}
\def\BDk {\ensuremath{\B^{\pm}\to\D K^\pm}\xspace}
\def\BDpi {\ensuremath{\B^{\pm}\to\D \pi^\pm}\xspace}
\def\BDpipi {\ensuremath{\Bz\to\D\pip\pim}\xspace}

\def\BDstKst {\ensuremath{ {\Bz\to\D^{\ast}\Kstarz}}\xspace}
\def\BsDstKst {\ensuremath{ {\Bsb\to\Dstarz\Kstarz}}\xspace}
\def\DKK {\ensuremath{\D\to\KS\Kp\Km}\xspace}
\def\Dpp {\ensuremath{\D\to\KS\pip\pim}\xspace}
\def\Dhh {\ensuremath{\D\to\KS h^+ h^-}\xspace}

\def\xm {\ensuremath{x_-}\xspace}
\def\ym {\ensuremath{y_-}\xspace}
\def\xp {\ensuremath{x_+}\xspace}
\def\yp {\ensuremath{y_+}\xspace}
\def\xpm {\ensuremath{x_\pm}\xspace}
\def\ypm {\ensuremath{y_\pm}\xspace}

\def\si {\ensuremath{s_i}\xspace}

\def\Fi {\ensuremath{F_i}\xspace}
\def\Fmi {\ensuremath{F_{-i}}\xspace}
\def\antiu {\ensuremath{\overline{\uquark}}\xspace}
\def\antic {\ensuremath{\overline{\cquark}}\xspace}
\def\mmsq {\ensuremath{m_-^2}\xspace}
\def\Db {\ensuremath{\bar{\D}}\xspace}
\def\DPcoords {\ensuremath{(\mmsq, \mpsq)}\xspace}
\def\AD {\ensuremath{A_\D\DPcoords}\xspace}
\def\ADb {\ensuremath{A_\Db\DPcoords}\xspace}
\def\mpsq {\ensuremath{m_+^2}\xspace}
\def\dzamp {\ensuremath{|\AD|}\xspace}

\def\dzbamp {\ensuremath{|\ADb|}\xspace}

\usepackage{hyperxmp}

\usepackage[pdftex,
            pdfauthor={\paperauthors},
            pdftitle={\paperasciititle},
            pdfkeywords={\paperkeywords},
            pdfcopyright={Copyright (C) \papercopyright},
            pdflicenseurl={\paperlicenceurl}]{hyperref}

\usepackage[bottom,flushmargin,hang,multiple]{footmisc}

\usepackage[all]{hypcap} 

\usepackage{xspace} 
\usepackage{upgreek}

\def\lhcb   {\mbox{LHCb}\xspace}

\def\babar  {\mbox{BaBar}\xspace}
\def\belle  {\mbox{Belle}\xspace}

\def\MagUp {\mbox{\em Mag\kern -0.05em Up}\xspace}

\ifthenelse{\boolean{uprightparticles}}%
{
 
 \def\Pgamma      {\ensuremath{\upgamma}\xspace}

 \def\Ppi         {\ensuremath{\uppi}\xspace}

 \def\PDelta      {\ensuremath{\Delta}\xspace}                 
 \def\PXi         {\ensuremath{\Xi}\xspace}                 
 \def\PLambda     {\ensuremath{\Lambda}\xspace}                 
 \def\PSigma      {\ensuremath{\Sigma}\xspace}                 
 \def\POmega      {\ensuremath{\Omega}\xspace}                 
 \def\PUpsilon    {\ensuremath{\Upsilon}\xspace}
 \let\oldPi\Pi
 \def\PPi         {\ensuremath{\oldPi}\xspace}

 \def\PB      {\ensuremath{\mathrm{B}}\xspace}                 
                  
 \def\PD      {\ensuremath{\mathrm{D}}\xspace}

 \def\PK      {\ensuremath{\mathrm{K}}\xspace}

 \def\Pb      {\ensuremath{\mathrm{b}}\xspace}                 
 \def\Pc      {\ensuremath{\mathrm{c}}\xspace}

 \def\Pi      {\ensuremath{\mathrm{i}}\xspace}

 \def\Ps      {\ensuremath{\mathrm{s}}\xspace}                 
                  
 \def\Pu      {\ensuremath{\mathrm{u}}\xspace}

 \def\thebaroffset{0.0em}
}
{
 
 \def\Pgamma      {\ensuremath{\gamma}\xspace}

 \def\Ppi         {\ensuremath{\pi}\xspace}

 \mathchardef\PDelta="7101
 \mathchardef\PXi="7104
 \mathchardef\PLambda="7103
 \mathchardef\PSigma="7106
 \mathchardef\POmega="710A
 \mathchardef\PUpsilon="7107
 \mathchardef\PPi="7105
                  
 \def\PB      {\ensuremath{B}\xspace}                 
                  
 \def\PD      {\ensuremath{D}\xspace}

 \def\PK      {\ensuremath{K}\xspace}

 \def\Pb      {\ensuremath{b}\xspace}                 
 \def\Pc      {\ensuremath{c}\xspace}

 \def\Pi      {\ensuremath{i}\xspace}

 \def\Ps      {\ensuremath{s}\xspace}                 
                  
 \def\Pu      {\ensuremath{u}\xspace}

 \def\thebaroffset{0.18em}
}
\newcommand{\offsetoverline}[2][\thebaroffset]{\kern #1\overline{\kern -#1 #2}}%

\makeatletter
\ifcase \@ptsize \relax
  \newcommand{\miniscule}{\@setfontsize\miniscule{4}{5}}
\or
  \newcommand{\miniscule}{\@setfontsize\miniscule{5}{6}}
\or
  \newcommand{\miniscule}{\@setfontsize\miniscule{5}{6}}
\fi
\makeatother

\DeclareRobustCommand{\optbar}[1]{\shortstack{{\miniscule (\rule[.5ex]{1.25em}{.18mm})}
  \\ [-.7ex] $#1$}}

\def\g      {{\ensuremath{\Pgamma}}\xspace}

\def\uquark    {{\ensuremath{\Pu}}\xspace}

\def\squark    {{\ensuremath{\Ps}}\xspace}

\def\cquark    {{\ensuremath{\Pc}}\xspace}

\def\bquark    {{\ensuremath{\Pb}}\xspace}

\def\pion   {{\ensuremath{\Ppi}}\xspace}
\def\piz    {{\ensuremath{\pion^0}}\xspace}
\def\pip    {{\ensuremath{\pion^+}}\xspace}
\def\pim    {{\ensuremath{\pion^-}}\xspace}

\def\kaon    {{\ensuremath{\PK}}\xspace}
\def\Kbar    {{\ensuremath{\offsetoverline{\PK}}}\xspace}

\def\KorKbar {\kern \thebaroffset\optbar{\kern -\thebaroffset \PK}{}\xspace}

\def\Kp      {{\ensuremath{\kaon^+}}\xspace}
\def\Km      {{\ensuremath{\kaon^-}}\xspace}

\def\KS      {{\ensuremath{\kaon^0_{\mathrm{S}}}}\xspace}

\def\Kstarz  {{\ensuremath{\kaon^{*0}}}\xspace}
\def\Kstarzb {{\ensuremath{\Kbar{}^{*0}}}\xspace}
\def\Kstar   {{\ensuremath{\kaon^*}}\xspace}

\def\Dbar    {{\ensuremath{\offsetoverline{\PD}}}\xspace}
\def\D       {{\ensuremath{\PD}}\xspace}
\def\Db      {{\ensuremath{\Dbar}}\xspace}
\def\DorDbar {\kern \thebaroffset\optbar{\kern -\thebaroffset \PD}\xspace}
\def\Dz      {{\ensuremath{\D^0}}\xspace}
\def\Dzb     {{\ensuremath{\Dbar{}^0}}\xspace}
\def\Dp      {{\ensuremath{\D^+}}\xspace}
\def\Dm      {{\ensuremath{\D^-}}\xspace}

\def\DpDm    {\ensuremath{\Dp {\kern -0.16em \Dm}}\xspace}
\def\Dstar   {{\ensuremath{\D^*}}\xspace}

\def\Dstarz  {{\ensuremath{\D^{*0}}}\xspace}

\def\B       {{\ensuremath{\PB}}\xspace}
\def\Bbar    {{\ensuremath{\offsetoverline{\PB}}}\xspace}

\def\BorBbar {\kern \thebaroffset\optbar{\kern -\thebaroffset \PB}\xspace}
\def\Bz      {{\ensuremath{\B^0}}\xspace}
\def\Bzb     {{\ensuremath{\Bbar{}^0}}\xspace}
\def\Bd      {{\ensuremath{\B^0}}\xspace}

\def\BdorBdbar {\kern \thebaroffset\optbar{\kern -\thebaroffset \Bd}\xspace}

\def\Bs      {{\ensuremath{\B^0_\squark}}\xspace}
\def\Bsb     {{\ensuremath{\Bbar{}^0_\squark}}\xspace}
\def\BsorBsbar {\kern \thebaroffset\optbar{\kern -\thebaroffset \Bs}\xspace}

\def\Y#1S{\ensuremath{\PUpsilon{(#1S)}}\xspace}

\def\LorLbar     {\kern \thebaroffset\optbar{\kern -\thebaroffset \PLambda}\xspace}

\def\to                 {\ensuremath{\rightarrow}\xspace}

\def\CP                {{\ensuremath{C\!P}}\xspace}

\def\AT#1     {\ensuremath{A_{\mathrm{T}}^{#1}}\xspace}           

\def\C#1      {\ensuremath{\mathcal{C}_{#1}}\xspace}                       
\def\Cp#1     {\ensuremath{\mathcal{C}_{#1}^{'}}\xspace}                    
\def\Ceff#1   {\ensuremath{\mathcal{C}_{#1}^{\mathrm{(eff)}}}\xspace}        
\def\Cpeff#1  {\ensuremath{\mathcal{C}_{#1}^{'\mathrm{(eff)}}}\xspace}       
\def\Ope#1    {\ensuremath{\mathcal{O}_{#1}}\xspace}                       
\def\Opep#1   {\ensuremath{\mathcal{O}_{#1}^{'}}\xspace}                    

\newcommand{\nospaceunit}[1]{\ensuremath{\text{#1}}}       
\newcommand{\aunit}[1]{\ensuremath{\text{\,#1}}}       

\newcommand{\tev}{\aunit{Te\kern -0.1em V}\xspace}
\newcommand{\gev}{\aunit{Ge\kern -0.1em V}\xspace}
\newcommand{\mev}{\aunit{Me\kern -0.1em V}\xspace}
\newcommand{\kev}{\aunit{ke\kern -0.1em V}\xspace}
\newcommand{\ev}{\aunit{e\kern -0.1em V}\xspace}
 
\newcommand{\mevc}{\ensuremath{\aunit{Me\kern -0.1em V\!/}c}\xspace}
\newcommand{\gevc}{\ensuremath{\aunit{Ge\kern -0.1em V\!/}c}\xspace}
\newcommand{\mevcc}{\ensuremath{\aunit{Me\kern -0.1em V\!/}c^2}\xspace}
\newcommand{\gevcc}{\ensuremath{\aunit{Ge\kern -0.1em V\!/}c^2}\xspace}

\def\mum  {\ensuremath{\,\upmu\nospaceunit{m}}\xspace}

\def\fb   {\ensuremath{\aunit{fb}}\xspace}
\def\invfb   {\ensuremath{\fb^{-1}}\xspace}

\newcommand{\chisq}{\ensuremath{\chi^2}\xspace}

\newcommand{\chisqip}{\ensuremath{\chi^2_{\text{IP}}}\xspace}

\def\gsim{{~\raise.15em\hbox{$>$}\kern-.85em
          \lower.35em\hbox{$\sim$}~}\xspace}
\def\lsim{{~\raise.15em\hbox{$<$}\kern-.85em
          \lower.35em\hbox{$\sim$}~}\xspace}

\def\pt         {\ensuremath{p_{\mathrm{T}}}\xspace}

\def\ptot       {\ensuremath{p}\xspace}

\def\degrees{\ensuremath{^{\circ}}\xspace}

\def\evtgen     {\mbox{\textsc{EvtGen}}\xspace}

\def\geant      {\mbox{\textsc{Geant4}}\xspace}

\def\photos     {\mbox{\textsc{Photos}}\xspace}

\def\pythia     {\mbox{\textsc{Pythia}}\xspace}

\def\tell1  {TELL1\xspace}
\def\ukl1   {UKL1\xspace}

\newcommand{\lhcborcid}[1]{\href{https://orcid.org/#1}{\hspace*{0.1em}\raisebox{-0.45ex}{\includegraphics[width=1em]{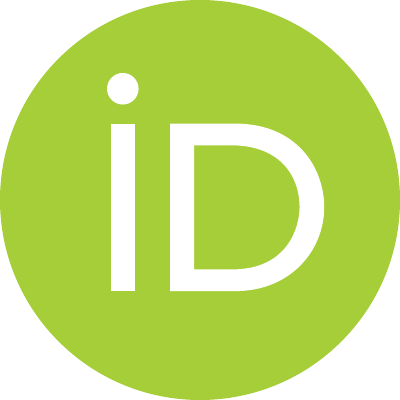}}}}

\usepackage{cite} 
\usepackage{mciteplus}

\usepackage{longtable} 

\begin{document}

\renewcommand{\thefootnote}{\fnsymbol{footnote}}
\setcounter{footnote}{1}


\begin{titlepage}
\pagenumbering{roman}

\vspace*{-1.5cm}
\centerline{\large EUROPEAN ORGANIZATION FOR NUCLEAR RESEARCH (CERN)}
\vspace*{1.5cm}
\noindent
\begin{tabular*}{\linewidth}{lc@{\extracolsep{\fill}}r@{\extracolsep{0pt}}}
\ifthenelse{\boolean{pdflatex}}
{\vspace*{-1.5cm}\mbox{\!\!\!\includegraphics[width=.14\textwidth]{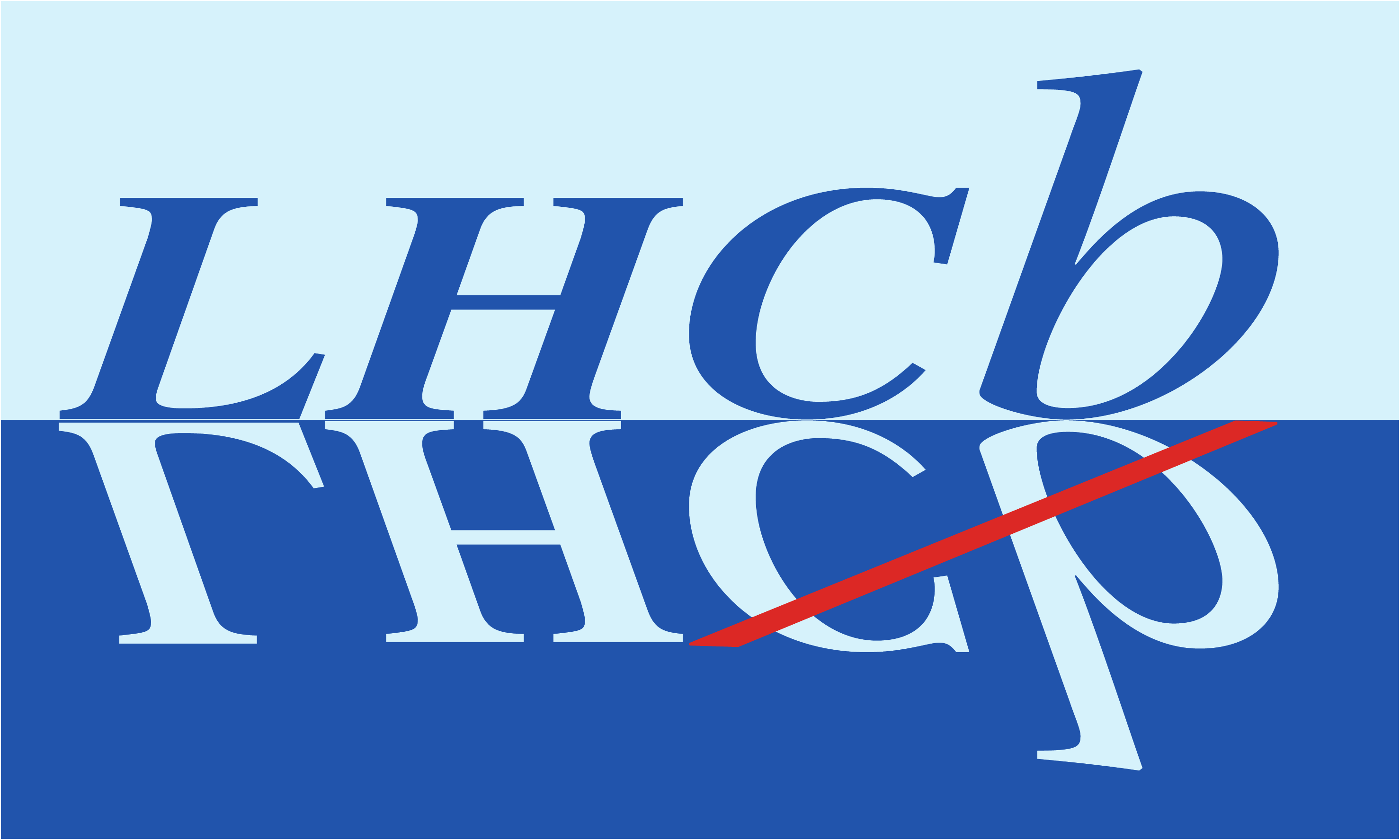}} & &}%
{\vspace*{-1.2cm}\mbox{\!\!\!\includegraphics[width=.12\textwidth]{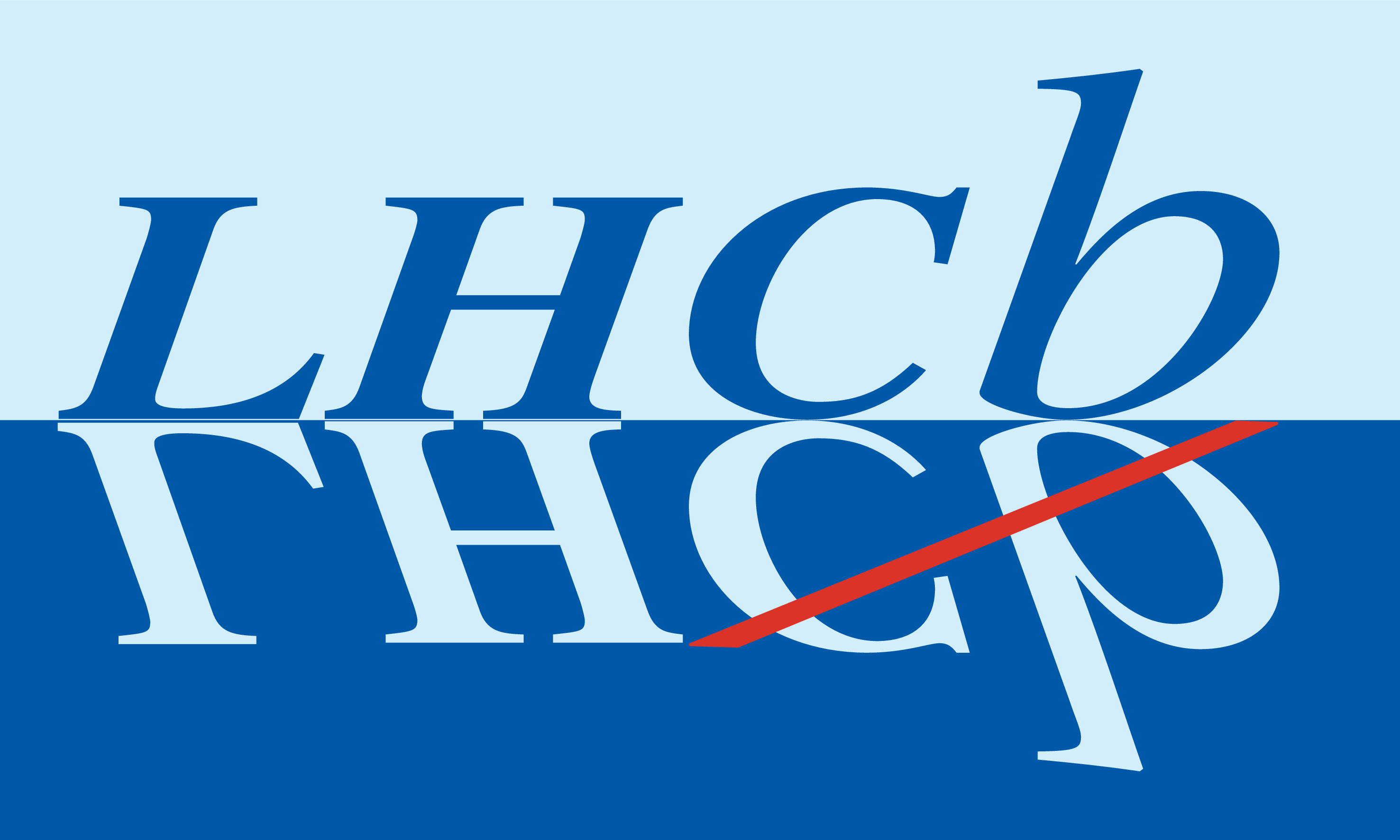}} & &}%
\\
 & & CERN-EP-2023-138 \\  
 & & LHCb-PAPER-2023-009 \\  
 & & \today \\ 
 & & \\
\end{tabular*}

\vspace*{4.0cm}

{\normalfont\bfseries\boldmath\huge
\begin{center}
\papertitle 
\end{center}
}

\vspace*{2.0cm}

\begin{center}
\paperauthors\footnote{Authors are listed at the end of this paper.}
\end{center}

\vspace{\fill}

\begin{abstract}
  \noindent
    A model-independent study of \CP violation in $\Bz\to\D\Kstarz$ decays is presented using data corresponding to an integrated luminosity of 9\invfb collected by the \lhcb experiment at centre-of-mass energies of $\sqrt{s}=7, \, 8$ and $13$\tev. The CKM angle $\gamma$ is determined by examining  the distributions of signal decays in phase-space bins of the self-conjugate $\D \to \KS h^+ h^-$ decays, where $h = \pi, K$.
    Observables related to \CP\ violation are measured and the angle $\gamma$ is determined to be $\gamma=(49^{+ 22}_{-19})\degrees$. Measurements of the amplitude ratio and strong-phase difference between the favoured and suppressed \Bz decays are also presented.
    
\end{abstract}

\vspace*{2.0cm}

\begin{center}
Published in Eur. Phys. J. C 84 (2024) 206
\end{center}

\vspace{\fill}

{\footnotesize 
\centerline{\copyright~\papercopyright. \href{\paperlicenceurl}{\paperlicence}.}}
\vspace*{2mm}

\end{titlepage}


\newpage
\setcounter{page}{2}
\mbox{~}
%
%
%
%


\renewcommand{\thefootnote}{\arabic{footnote}}
\setcounter{footnote}{0}

\cleardoublepage


\pagestyle{plain} 
\setcounter{page}{1}
\pagenumbering{arabic}


\section{Introduction}
\label{sec:intro}
In the Standard Model (SM), the Cabibbo–Kobayashi–Maskawa (CKM) matrix~\cite{Cabibbo:1963yz, Kobayashi:1973fv} describes flavour-changing weak transitions of quarks.
The phase difference between the CKM matrix elements for \mbox{$\bquark\to\uquark$} and \mbox{$\bquark\to\cquark$} quark transitions, defined as \mbox{$\gamma \equiv \rm{arg}\left(-V_{\textit{ud}}^{\phantom{\ast}}V_{\textit{ub}}^{\ast}/V_{\textit{cd}}^{\phantom{\ast}}V_{\textit{cb}}^{\ast}\right)$}, is of particular interest because it is measurable in purely tree-level decays and has negligible theoretical uncertainty~\cite{Brod:2013sga}.
Therefore, the SM can be tested by comparing direct measurements of $\gamma$ with indirect determinations obtained by fitting the CKM unitarity triangle.
The average value of the direct measurements is \mbox{$\gamma_{\rm{direct}}=(66.2^{+3.4}_{-3.6})\degrees$}~\cite{HFLAV21}, which agrees at current precision with the indirectly determined value \mbox{$\gamma_{\rm{indirect}}=(65.6^{+0.9}_{-2.7})\degrees$}~\cite{CKMfitter2015} or \mbox{$\gamma_{\rm{indirect}}=(65.8 \pm 2.2)\degrees$}~\cite{UTfit:2022hsi} depending on the statistical approach used.
A more stringent test requires improving the precision on both the direct and indirect determinations of $\gamma$.

\begin{figure}[b]
    \centering
    \setlength{\tabcolsep}{0pt}
        \includegraphics[width=0.49\textwidth]{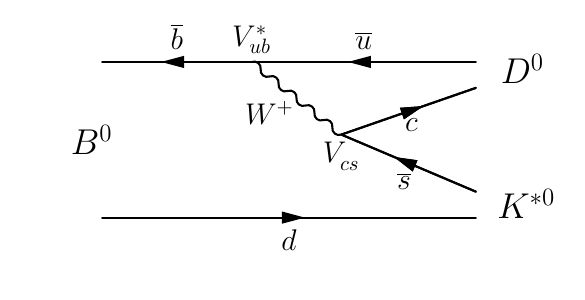}
        \includegraphics[width=0.49\textwidth]{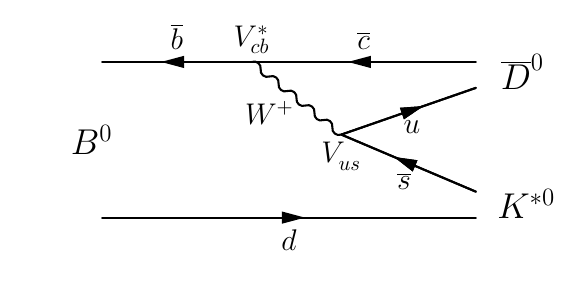}
    \caption{Feynman diagrams for the (left) suppressed and (right) favoured \BDKst decays.}
\label{fig:feynman}
\end{figure}

The precision on $\gamma$ is dominated by the measured \CP violation in the interference between \mbox{$\bquark\to\uquark\antic\squark$} and \mbox{$\bquark\to\cquark\antiu\squark$} quark transitions in  \mbox{$\B^\pm \to \D K^\pm$} decays. Here, \D represents a superposition of \Dz and \Dzb mesons. However it is possible to gain complementary information from the \mbox{$\Bz \to \D\Kstar(892)^{0}$} decay\footnote{The inclusion of charge-conjugate processes is implied, unless explicitly stated otherwise.}. While this decay has a lower branching fraction compared to the \mbox{$\B^\pm \to \D K^\pm$} channel, the interference between the favoured and suppressed \BDKst\ decays is expected to be a factor of 3 larger since both amplitudes are colour suppressed, leading to a higher per-event sensitivity to $\gamma$. Feynman diagrams of the two possible $\Bz$ decays are shown in Fig.~\ref{fig:feynman}. The flavour of the \B meson at the point of decay is unambiguously provided by the charge of the kaon from the \mbox{$\Kstar(892)^{0}  \to \Kp\pim$} decay, and hence the analysis of this channel can proceed without considering time dependence. Interference between the two amplitudes is accessed through reconstruction of the $\D$ meson in final states common to both \Dz and \Dzb. For the analysis presented here the \D mesons are reconstructed in the self-conjugate \mbox{\Dhh} decay modes ($h = \pi, K$). The Belle~\cite{Belle:2012bcb, Belle:2015roy} and BaBar~\cite{PhysRevD.80.031102} collaborations have used the \mbox{\BDKst} channel to determine $\gamma$ with various final states of \D decay, including \Dpp.
However, the most precise measurements using the \BDKst decay mode have been made by the \lhcb experiment~\cite{LHCb-PAPER-2019-021,LHCb-PAPER-2016-006}.

The work presented here uses data collected with the \lhcb detector in proton-proton ($pp$) collisions at centre-of-mass energies of $\sqrt{s}=7, \, 8$ and $13$\tev between 2011--2012 and 2015--2018, corresponding to an integrated luminosity of 9\invfb.
The experimental procedure employed here closely follows that described in Ref.~\cite{LHCb-PAPER-2016-006}, where \CP violation observables that are related to $\gamma$ are determined through the distributions of  \mbox{\BDKst} and \mbox{$\Bzb\to\D\Kstarzb$} decays in regions of the \mbox{\Dhh} decay phase space~\cite{ggsz1, Bondar:2005ki, Bondar:2008hh, Giri:2003ty}. The extraction of $\gamma$ requires knowledge of the \D decay strong-phase parameters, which were directly determined by the BESIII~\cite{BESIII:2020khq,BESIII:2020hpo,BESIII:2020hlg} and CLEO~\cite{CLEO:2010iul} collaborations. Therefore, the measurement avoids using any \D decay amplitude model, thus is free of any systematic uncertainty attributed to such models.

The data set used for the work presented here is increased compared to Ref.~\cite{LHCb-PAPER-2016-006}.
In addition, a number of procedural improvements are made, such as adopting a more optimal division of \Dpp phase space, and employing an improved strategy to handle the varying reconstruction efficiency over \D decay phase space. Furthermore, the strong-phase inputs are updated to reflect the most recent combination of results from CLEO and BESIII~\cite{BESIII:2020khq, BESIII:2020hpo}.

\section{Analysis overview}
The amplitudes of the favoured and suppressed $\Bz \to \D\Kp\pim$ decays, where the $\Kp\pim$ is not restricted to the \Kstarz resonance, can be written as
\begin{align}
    A(\Bz \to \Dzb \Kp\pim) \equiv &\ A_c(p) e^{i\delta_c(p)} &\ \textrm{(favoured)}, \\
    A(\Bz \to \Dz \Kp\pim) \equiv &\ A_u(p) e^{i\left[\delta_u(p) + \gamma\right]} &\ \textrm{(suppressed)},
\end{align}
where $A_{c(u)}$ and $\delta_{c(u)}$ are the magnitude and strong-phase of the decay corresponding to the \mbox{$\bquark\to\cquark (\uquark)$} transitions, respectively, and $p$ is the phase-space coordinate of the $D\Kp\pim$ final state. 
The equivalent amplitudes for the \CP conjugate, \mbox{$\Bzb\to\D\Km\pip$}, are given by transforming $\gamma \to -\gamma$.
In this analysis, the amplitude ratio (\rB) and strong-phase difference (\deltaB) between the favoured and suppressed signal decays are
measured alongside the angle $\gamma$. They are defined as
\begin{align}
    r_\Bz^2 &= \frac{\int_{\Kstarz} \mathrm{d}p\, A_u(p)^2}{\int_{\Kstarz} \mathrm{d}p\, A_c(p)^2},\\[4pt]
    \kappa e^{i\delta_\Bz} &= \frac{\int_{\Kstarz} \mathrm{d}p\, A_c(p) A_u(p) e^{i\left[\delta_u(p) - \delta_c(p)\right]}}{\sqrt{\int_{\Kstarz} \mathrm{d}p\, A_c(p)^2} \sqrt{\int_{\Kstarz} \mathrm{d}p\, A_u(p)^2}},
\end{align}
where the integral is performed over the \mbox{\BDKst} region of the \mbox{$\Bz\to\D\Kp\pim$} phase space.
The coherence factor, $\kappa$, accounts for pollution from decays that are not \BDKst, and satisfies $0 \leq \kappa \leq 1$.
The value of $\kappa=0.958^{+0.005}_{-0.046}$ is used as a direct input from the \lhcb amplitude analysis of $\Bz\to\D\Kp\pim$ decays described in Ref.~\cite{LHCb-PAPER-2015-059}.
The kinematic selection of the \Kstarz candidates used in this work follows that of Ref.~\cite{LHCb-PAPER-2015-059} to match the phase-space region in which $\kappa$ is evaluated. 

\begin{figure}[t]
    \centering
    \setlength{\tabcolsep}{0pt}
        \includegraphics[width=0.48\textwidth]{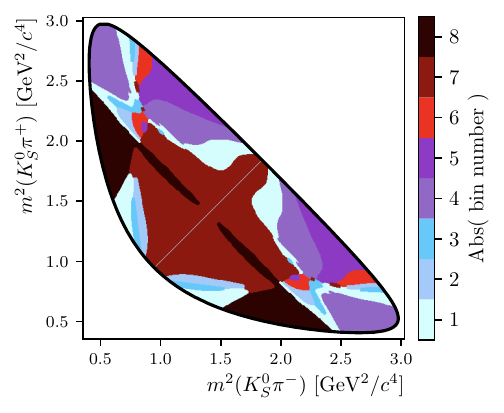}
        \includegraphics[width=0.48\textwidth]{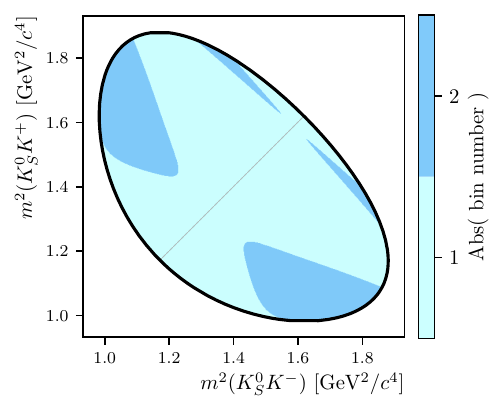}
    \caption{Dalitz plot binning schemes used for (left) \mbox{\Dpp} and (right) \mbox{\DKK} decays.}
\label{fig:binning_schemes}
\end{figure}

The amplitudes for the $\Dz\to\KS h^+ h^-$ and $\Dzb\to\KS h^+ h^-$ decays are written as \mbox{$\AD = |\AD|e^{i\delta(m_-^2,\ m_+^2)}$}  and \mbox{$\ADb = |\ADb|e^{i\overline{\delta}(m_-^2,\ m_+^2)}$}, respectively, where $m_\pm^2 = m^2(\KS h^\pm)$ are the Dalitz plot coordinates.
The \D decay phase space is divided into independent regions.
A scheme is used with $2\mathcal{N}$ bins labelled from $i=-\mathcal{N}$ to $i=\mathcal{N}$ (excluding 0).
The division is symmetrical about the line $\mmsq = \mpsq$, and a bin where $\mmsq > \mpsq$ ($\mmsq < \mpsq$) is referred to as the $i^{th}$ ($-i^{th}$) bin.
The `optimal' ~\cite{CLEO:2010iul} (`2-bin') scheme with $\mathcal{N}=8$ ($\mathcal{N}=2$) bins is used for the \mbox{\Dpp} (\mbox{\DKK}) mode, and is displayed in Fig.~\ref{fig:binning_schemes}.

The total amplitude of the \BKshhKpi decay is given by
\begin{align}
A(\BKshhKpi) =\ &A_c(p) e^{i\delta_c(p)} \ADb \\ \nonumber &+ A_u(p) e^{i\left[\delta_u(p) + \gamma\right]} \AD,
\end{align}
where that of the \CP conjugate \Bzb decay is found by transforming $\gamma \to -\gamma$ and $m_- \leftrightarrow m_+$.
Squaring the total amplitude and integrating over the \Kstarz phase-space region gives
\begin{align}
\Gamma(\Bz\to (\KS h^+ h^-)_\D \Kstarz) \propto\ &|A_\Db|^2 + \rB^2 |A_\D|^2 \\ \nonumber
&+ 2\kappa\rB |A_\D| |A_\Db| \left[\cos(\deltaB + \gamma)\cos\delta_\D - \sin(\deltaB + \gamma)\sin\delta_\D \right],
\end{align}
where the Dalitz plot coordinates of the \D decay strong-phase difference, defined as \mbox{$\delta_D(m_-^2,\ m_+^2) = \delta(m_-^2,\ m_+^2) - \overline{\delta}(m_-^2,\ m_+^2)$}, and magnitudes have been omitted for brevity.
The expression for the decay rate integrated over a Dalitz plot bin is given by
\begin{align}
\Gamma_{i}(\Bz \to (\KS h^+ h^-)_\D\Kstarz) \propto\ & K_{-i} + \rB^2 K_{i} \\ \nonumber & + 2\kappa \rB \sqrt{K_i K_{-i}}[\cos(\deltaB + \gamma) c_{i} - \sin(\deltaB + \gamma) s_{i}],
\end{align}
where the \D decay magnitude and strong-phase difference have been replaced by integrals over Dalitz plot bins
\begin{align}
    K_i &= \int_i \mathrm{d}\mmsq \mathrm{d}\mpsq\, \dzamp^2,  \\
    c_i &= \frac{1}{\sqrt{K_i K_{-i}}}\int_{i} \mathrm{d}\mmsq \mathrm{d}\mpsq\, \dzamp\dzbamp
    \cos\delta_D(\mmsq,\mpsq),\\
    s_i &= \frac{1}{\sqrt{K_i K_{-i}}}\int_{i} \mathrm{d}\mmsq \mathrm{d}\mpsq\, \dzamp\dzbamp \sin\delta_D(\mmsq,\mpsq).
\end{align}
Swapping the coordinates $m_- \leftrightarrow m_+$ is equivalent to a bin transformation $i \leftrightarrow -i$, and results in the relations $c_i = c_{-i}$ and $s_i = - s_{-i}$.

Experimentally, candidate yields are determined instead of the decay rates. 
Detector, reconstruction and selection related efficiencies are accounted for by using a set of parameters referred to as $F_i$ that are determined in each bin. 
They are defined as
\begin{equation}
    F_{i} \equiv \frac{\int_{i} \mathrm{d}\mmsq \mathrm{d}\mpsq\, \dzamp^2 \eta(\mmsq, \mpsq)}{\sum_j \int_j \mathrm{d}\mmsq \mathrm{d}\mpsq\, \dzamp^2\eta(\mmsq, \mpsq)},
    \label{eq:Fi_values}
\end{equation}
where $\eta(\mmsq, \mpsq)$ is the efficiency profile which varies over the \D decay phase space. 
The $F_i$ are the efficiency-modulated $K_i$ parameters, and are dependent on the experimental resolution and selection efficiency.
A similar efficiency adjustment is not included in the $c_i$ and $s_i$ parameters because the effect is small, however a systematic uncertainty is included to account for this assumption.
The \Fi parameters have been determined using \mbox{\BDpi} decays~\cite{LHCb-PAPER-2020-019}. As these parameters are selection dependent, they are only valid for use in this analysis under the assumption that the relative variation in $\eta(\mmsq,\mpsq)$ between \D meson decays in \mbox{\BDpi} and \mbox{\BDKst} is the same.
Differences in the efficiency profiles are minimised by employing a similar selection between the \BDKst and \BDpi decays and small residual differences are determined using simulation samples and used to assign systematic uncertainties on the \CP violation observables.
The yields of \Bz and \Bzb decays in a Dalitz plot bin are given by
\begin{align}
    N_i(\Bz) = h^\Bz \left[ F_{-i} + (\xp^2 + \yp^2)F_{i} + 2\kappa\sqrt{F_{i}F_{-i}} (\xp c_{i} - \yp s_{i})\right], \label{eq:N_Bz} \\
    N_i(\Bzb) = h^\Bzb \left[ F_{i} + (\xm^2 + \ym^2)F_{-i} + 2\kappa\sqrt{F_{i}F_{-i}} (\xm c_{i} + \ym s_{i})\right], \label{eq:N_Bzb}
\end{align}
where the \CP violation observables~\cite{BaBar:2005rek}, \xpm and \ypm,  are related to the physics parameters by
\begin{align}
    \xpm \equiv \rB \cos(\deltaB \pm \gamma), \\
    \ypm \equiv \rB \sin(\deltaB \pm \gamma).
\end{align}
These observables have improved statistical behaviour in comparison to determining $\gamma$, $\rB$ and $\deltaB$ directly. The two normalisation constants in Eqs.~(\ref{eq:N_Bz}) and~(\ref{eq:N_Bzb}), $h^\Bz$ and $h^\Bzb$, are the observed total yields of the \Bz and \Bzb decay modes. The use of two separate normalization constants is intentional, as nearly all detector and production asymmetries are absorbed into these parameters leaving the measurement insensitive to these effects. Equations~(\ref{eq:N_Bz}) and~(\ref{eq:N_Bzb}) are used to fit the data and determine the \CP violation observables. In the fit, the external input parameters $\kappa$~\cite{LHCb-PAPER-2015-059}, $F_i$~\cite{LHCb-PAPER-2020-019}, 
$c_i$ and $s_i$~\cite{BESIII:2020khq, BESIII:2020hpo} are fixed to their measured central values.

The \mbox{$\Bsb\to\Dz\Kstarz$} decay, which has identical final state particles is reconstructed alongside the signal channel. 
In principle, the method described in this section could also be applied to \mbox{$\Bsb\to\D\Kstarz$} decays. However, the sensitivity to $\gamma$ is significantly lower due to reduced interference between the two final state paths. The values of the CKM elements~\cite{PDG2022} can be used to predict that the ratio of suppressed to favoured amplitudes, $r_{\Bs}\sim0.02$, is over a factor of 10 less than in \mbox{\BDKst} decays~\cite{LHCb-PAPER-2021-033}. In this analysis it is assumed that the \CP violation in the \mbox{$\Bsb\to\D\Kstarz$} decay is zero and it is not treated as a signal decay mode. Thus in the remainder of the paper this decay is referred to as the \mbox{\BsDKst} background with a flavour specific $D$ meson.

\section{Detector and simulation}
The \lhcb detector~\cite{LHCb-DP-2008-001,LHCb-DP-2014-002} is a single-arm forward spectrometer covering the \mbox{pseudorapidity} range $2<\eta <5$, designed for the study of particles containing \bquark or \cquark quarks. 
The detector includes a high-precision tracking system consisting of a silicon-strip vertex detector surrounding the $pp$ interaction region, a large-area silicon-strip detector located upstream of a dipole magnet with a bending power of about $4{\mathrm{\,Tm}}$, and three stations of silicon-strip detectors and straw drift tubes placed downstream of the magnet.
The tracking system provides a measurement of the momentum, \ptot, of charged particles with a relative uncertainty that varies from 0.5\% at low momentum to 1.0\% at 200\gevc. 
The minimum distance of a track to a primary $pp$ collision vertex (PV), the impact parameter (IP), is measured with a resolution of $(15+29/\pt)\mum$, where \pt is the component of the momentum transverse to the beam, in\,\gevc. 
Different types of charged hadrons are distinguished using information from two ring-imaging Cherenkov detectors. 
Photons, electrons and hadrons are identified by a calorimeter system consisting of scintillating-pad and preshower detectors, an electromagnetic 
and a hadronic calorimeter. 
Muons are identified by a system composed of alternating layers of iron and multiwire proportional chambers.

The online event selection is performed by a trigger, which consists of a hardware stage, based on information from the calorimeter and muon systems, followed by a software stage, which applies a full event reconstruction.
The events that are selected for the analysis either have final-state tracks of the signal decay that are subsequently associated with an energy deposit in the calorimeter system that satisfies the hardware stage trigger, or are selected because one of the other particles in the event, not reconstructed as part of the signal candidate, fulfils any hardware stage trigger requirement. 
At the software stage, it is required that at least one particle should have high \pt and high \chisqip, where \chisqip is defined as the difference in the primary vertex fit \chisq with and without the inclusion of that particle. 
A multivariate algorithm~\cite{BBDT} is used to select secondary vertices consistent with being a two-, three-, or four-track b-hadron decay. 

Simulated data are required to determine the invariant-mass shapes of signal and background components, and to compute relative selection efficiencies.
In the simulation, $pp$ collisions are generated using \pythia~\cite{Sjostrand:2007gs,*Sjostrand:2006za} with a specific \lhcb configuration~\cite{LHCb-PROC-2010-056}.
Decays of unstable particles are described by \evtgen~\cite{Lange:2001uf}, in which final-state radiation is generated using \photos~\cite{davidson2015photos}.
The decays \mbox{\Dpp} and \mbox{\DKK} are generated uniformly over phase space.
The interaction of the generated particles with the detector, and its response, are implemented using the \geant toolkit~\cite{Allison:2006ve, *Agostinelli:2002hh} as described in Ref.~\cite{LHCb-PROC-2011-006}.

\section{Candidate selection}
\label{sec:selection}
All tracks and decay vertices are required to be of good quality, and the reconstructed  mass of the \KS, \D and \Kstarz candidates must be close to their known values~\cite{PDG2022}.
The \KS candidates are formed from two oppositely charged pions, where the tracks are reconstructed using hits in the vertex detector and other downstream tracking stations, or only the latter. 
These track types are referred to as \textit{long} and \textit{downstream}, respectively, and are treated separately since the former leads to better mass, momentum and vertex resolution on the $\KS$ candidate and higher reconstruction efficiency. 
A \D meson candidate is formed by combining a \KS candidate with two oppositely charged pions. Particle identification (PID) requirements are placed on the particles, to reduce background from $\D\to\KS\Kp\pim$ decays, semileptonic \D decays, and hadronic decays in flight to leptons. 
A requirement is placed on the displacement of the \D meson vertex from the \B meson vertex to reduce background from \B decays to the final state particles without the intermediate \D meson.
The \D meson candidate is then combined with a \Kstarz candidate, which is formed by combining a pion and kaon, with strict PID requirements to suppress $\Bz\to\D\pip\pim$ backgrounds and thus allow for correct identification of the \B-meson flavour.
A criterion is applied on the \Kstarz helicity angle, $\theta^\ast$, defined as the angle between the kaon from the \Kstarz decay and the opposite of the \B momentum in the \Kstarz rest frame, to exploit differences in the angular distributions of the signal and background candidates.
In signal decays, a \B meson decays to a vector and pseudo-scalar final state, so the corresponding distribution of $|\cos\theta^\ast|$ peaks at 1, whereas it is flat for background candidates formed from random combinations of tracks, referred to as combinatorial background.
Therefore, candidates are rejected if the value of $|\cos\theta^\ast|$ is below a threshold that is chosen to match that applied in Ref.~\cite{LHCb-PAPER-2015-059}.

A kinematic fit is performed to improve the resolution of the invariant-mass of the \Bz candidates and Dalitz plot coordinates. 
In this fit, the masses of the \D and \KS candidates are constrained to their known values~\cite{PDG2022}, and the momentum of the \Bz meson is required to be parallel to the vector linking the \Bz\ decay vertex and the associated PV, which is defined as the PV leading to the smallest IP of the \Bz candidate.

A boosted decision tree (BDT) classifier~\cite{Breiman,AdaBoost} is employed to reduce combinatorial background. 
It is trained on \mbox{\BDKst} decays with \mbox{\Dpp} separately for candidates with \textit{long} and \textit{downstream} \KS track types, and is applied to both \D decay modes. 
Signal is represented by simulated decays, and combinatorial background is represented by \mbox{\BDKst} candidates in data with an invariant mass between 5800 and 6200\mevcc. 
The set of input variables are predominantly based on the decay topology and kinematics. They are taken from the BDT classifier applied in the analysis of \BDh decays outlined in Ref.~\cite{LHCb-PAPER-2020-019}. Since there is an extra track in \mbox{\BDKst} decays, the \ptot, \pt and \chisqip of the pion from the \Kstarz decay are also included.
The optimal BDT classifier selection criterion is chosen to minimise the statistical uncertainty on $\gamma$ and is determined with pseudoexperiments.

\begin{figure}[t]
    \centering
        \includegraphics[width=0.47\textwidth]{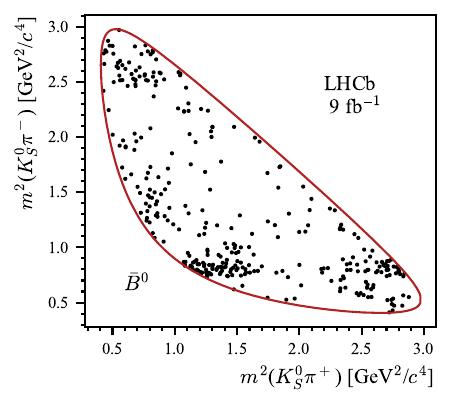}
        \includegraphics[width=0.47\textwidth]{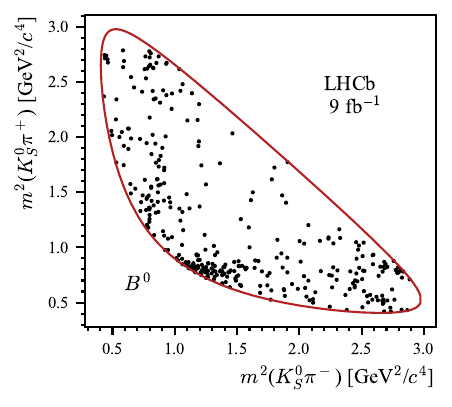}
        \includegraphics[width=0.47\textwidth]{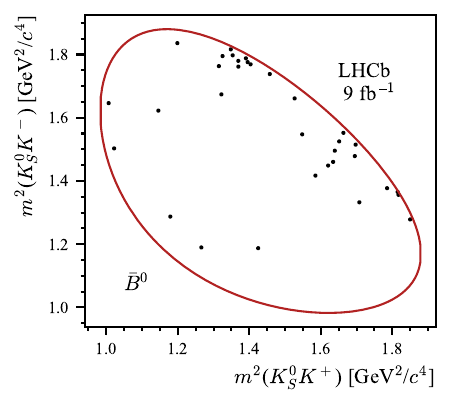}
        \includegraphics[width=0.47\textwidth]{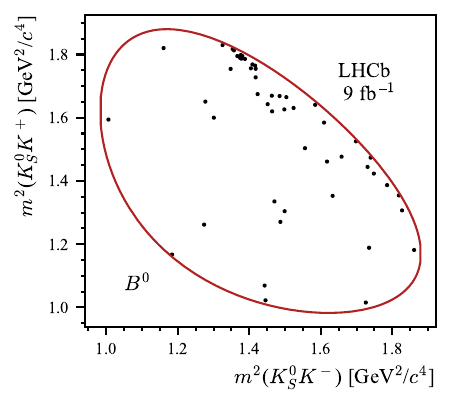}
    \caption{Dalitz plots of selected candidates for (left) \Bzb and (right) \Bz decays followed by the (upper) \mbox{\Dpp} and (lower) \mbox{\DKK} decay. Candidates that have an invariant mass within a 30 \mevcc region either side of the \Bz mass are displayed. The kinematic boundaries are plotted as continuous red solid curves.}
\label{fig:dalitz_selected}
\end{figure}

Figure~\ref{fig:dalitz_selected} displays Dalitz plot distributions of fully selected \mbox{\BDKst} candidates that have an invariant mass within $\pm30$ \mevcc of the \Bz mass~\cite{PDG2022}, where the signal purity is approximately 60\%.
They are displayed in four categories given by the \D decay and \B-meson flavour, and candidates from both \KS track types are combined for visualisation purposes only.

\section{Fit to determine the \texorpdfstring{\textbf{\CP}}{CP} violation observables}
\label{sec:fit}
A two-stage fit strategy is adopted to determine the \CP violation observables. 
The same model is used for both stages in an unbinned maximum likelihood fit to the invariant-mass distribution of \mbox{\BDKst} candidates in the region 5200--5800\mevcc. The lower end of the fit range is chosen to remove background from \CP violating $\Bz \to \Dstar \Kstar$ decays.
The first stage, referred to as the global fit, is used to understand the background composition and parameterize the invariant-mass distribution. The candidates in this fit are divided into four groups, given by the \D decay mode and the \KS track type. In the second stage the data are simultaneously fitted across 80 categories given by the \D decay mode, \KS track type, \B-meson flavour and Dalitz plot bin.

Due to the similarities in the final state, the signal and \mbox{\BsDKst} decays have a similar invariant mass shape. Both are modelled by a function with a Gaussian core and asymmetric tails,
\begin{small}
\begin{center}
    \begin{equation}
    \label{eq:crj}
        f(m) = \begin{cases}
        \exp(-\delta m^2 (\frac{1+\beta \delta m^2}{f_L})), ~~ \delta m<0 \\
        \exp(-\delta m^2 (\frac{1+\beta \delta m^2}{f_R})), ~~ \delta m>0 \\
        \end{cases}
\text{where \ }
        \begin{gathered}
            \delta m = m-\mu, \\
            f_L = 2\sigma_L^2 + \alpha_L \delta m^2, \\ 
            f_R = 2(\frac{\sigma_L}{r})^2 + \alpha_R \delta m^2 \\
        \end{gathered}
    \end{equation}
\end{center}
\end{small}
where $\mu$ is the mean, $\beta$ is the asymmetry, $\sigma_{L,R}$ and $\alpha_{L,R}$ describe the left and right widths and tails, respectively. 
The $\beta$, $\alpha$ and width ratio, $r=\frac{\sigma_L}{\sigma_R}$, parameters are fixed to values determined from simulation.
The mean of the distribution representing \mbox{\BsDKst} candidates is a free parameter shared between the categories, whilst that of signal is constrained using the known mass difference, \mbox{$m(\Bs) - m(\Bz) = (87.42 \pm 0.16)$\mevcc }\cite{PDG2022}. 
Finally, the width is shared between signal and \mbox{\BsDKst} decays for both \D decay modes but different for \textit{long} and \textit{downstream} \KS track categories.

The dominant physics background near the signal is from $\Bsb\to\Dstarz\Kstarz$ candidates with the $\Dstarz$ decaying to a $\Dz$ and an unreconstructed \g or \piz. 
The mass model of this background is described by four components depending on which particle is missed and whether the helicity state of the \Dstarz is 0 or $\pm1$ (the distributions of the $\pm1$ states are indistinguishable). 
The parameters describing the shape of each component are fixed to the values determined in simulation. 
It is not possible to determine the relative fractions of these four components reliably using data collected with the self-conjugate \Dhh modes, because the invariant-mass region below 5200 \mevcc is dominated by a mix of \BDstKst and \BsDstKst decays and their distributions significantly overlap. 
However, in Cabibbo-favoured \D meson decays the low invariant-mass region is dominated by either \Bsb or \Bz decays. 
This advantage is used by fitting the invariant mass distribution of candidates reconstructed as \mbox{$\Bsb\to \Dz(\to K^-\pi^+)\Kstarz$} decays to determine the relative fractions of each partially reconstructed \mbox{$\Bsb\to\Dstarz\Kstarz$} component. The selection of candidates and the mass fit parameterisation follows that described in Ref.~\cite{LHCb-PAPER-2019-021}, but the data set is increased to include that collected in 2017 and 2018. Given the studies in Ref.~\cite{LHCb-PAPER-2021-043}, contamination from $\Bsb \to \Dstarz K \pi$ decays that do not include the \Kstarz resonance is small and this background will be subsumed into either the \BsDstKst shapes or the combinatorial background. 
A small amount of \BDstKst decays leaks into the fit range. Their invariant-mass shape and yield ratio are determined in a similar way to that for the \BsDstKst background by studying candidates reconstructed as $\Bz\to\D(\to K\pi)\Kstarz$.

Backgrounds from ${\B^\pm \to \D K^\pm}$ decays plus a random pion, and misidentified ${\Bz\to\D\pip\pim}$ decays are represented by shapes determined using simulation samples.
The relative yield of both are fixed with respect to that of the \mbox{\BsDKst} candidates.
The ratio of \mbox{${\B^\pm \to \D K^\pm}$} decays is determined using branching fractions, fragmentation fractions~\cite{LHCb-PAPER-2012-037,LHCb-PAPER-2020-046} and selection efficiencies in simulation, where differences from data are determined to be negligible.
The ratio for misidentified ${\Bz\to\D\pip\pim}$ is determined from the results of fits to $\Bz\to (\D \to K\pi)\Kstarz$ decays.
Finally, the combinatorial background is described by an exponential function, where the yield and slope are freely varying parameters in each category.

\begin{figure}[tb]
    \centering
        \includegraphics[width=0.49\textwidth]{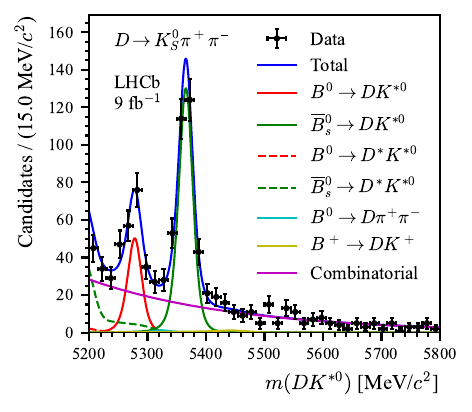}
        \includegraphics[width=0.49\textwidth]{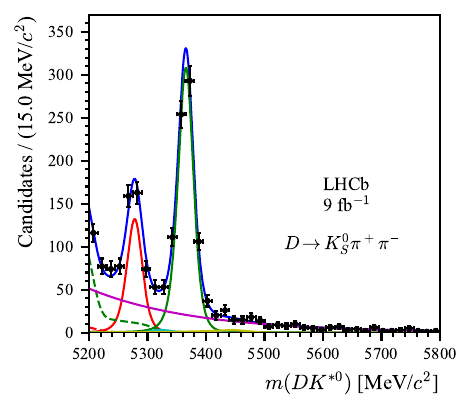}
        \includegraphics[width=0.49\textwidth]{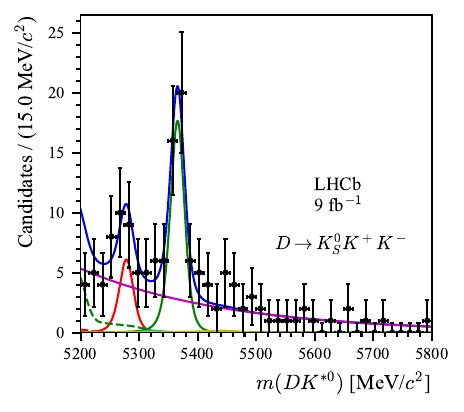}
        \includegraphics[width=0.49\textwidth]{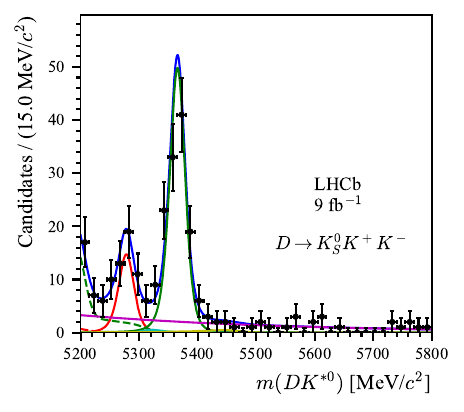}
    \caption{Invariant-mass distributions of \mbox{\BDKst} candidates with (upper) \mbox{\Dpp} and (lower) \mbox{\DKK} decays, separated by the (left) \textit{long} and (right) \textit{downstream} \KS track type. The data are overlaid with the global fit projection. }
\label{fig:global_fits}
\end{figure}

The projections of the global fits are displayed in Fig.~\ref{fig:global_fits}.
Table~\ref{tab:global_fit_yields} details the yields of each component in a 30\mevcc region either side of the \Bz mass~\cite{PDG2022}.  The total signal yield and purity are $434 \pm 32$ and $(57 \pm 5)\%$, respectively. The dominant backgrounds in the signal region are from combinatorial candidates and \BsDstKst decays. Other sources are negligible in comparison.

\begin{table}[tb]
\caption{Yield of each component in a 30\mevcc region either side of the \Bz mass as determined by the global fit in four categories. Yields are either determined directly or through a combination of fit parameters. The uncertainties are determined through propagation and further modulated by integration within the region.
Some backgrounds have negligible yields in the aforementioned invariant-mass region.}
\begin{adjustbox}{width=1\textwidth}
    \centering
    \begin{tabular}{ccccc}
    \toprule
         Component & \mbox{\Dpp}          & \mbox{\Dpp}                & \mbox{\DKK}          & \mbox{\DKK} \\
                   & \textit{long} & \textit{downstream} & \textit{long} & \textit{downstream} \\ 
        \midrule
        \BDKst             & {102} $\pm$ {17} & {288} $\pm$ {25} & {12} $\pm$ {6} & {32} $\pm$ {8} \\
        \BsDKst            & \phantom{0}{2.4} $\pm$ {0.4} & \phantom{0}{7.1} $\pm$ {0.6} & \phantom{0}{0.32} $\pm$ {0.08} & \phantom{0}{1.2} $\pm$ {0.2} \\
     Combinatorial      & {84} $\pm$ {8} & {133} $\pm$ {11} & {16} $\pm$ {3} & {11} $\pm$ {4} \\
     \BsDstKst          & {17.1} $\pm$ {1.4} & {44} $\pm$ {2} & \phantom{0}{2.3} $\pm$ {0.5} & \phantom{0}{7.1} $\pm$ {0.8} \\
        \BDstKst           & $\leq 1$& $\leq 1$& $\leq 1 $& $\leq 1$\\
        
        $\Bz\to\D\pip\pim$ & $\leq 1 $& \phantom{0}{1.8} $\pm$ {0.5} & $\leq 1$& $\leq 1 $\\
        $\B^\pm\to\D K^\pm$& $\leq1$ & \phantom{0}{2.0} $\pm$ {0.4} & $\leq 1$ & $\leq 1$ \\
       
        \bottomrule
    \end{tabular}
    \end{adjustbox}
    \label{tab:global_fit_yields}
\end{table}

Simulation is used to verify that the component shapes do not vary across the Dalitz plot.
Therefore, the same model is applied in the fit to extract the \CP violation observables as for the global fit.
The yield of each component, excluding combinatorial background, in a Dalitz plot bin is parameterised by the integrated yield multiplied by the expected fraction of candidates in that bin.
For example, the distribution of signal candidates is described by Eqs.~(\ref{eq:N_Bz}) and (\ref{eq:N_Bzb}) where $h_\Bz$ and $h_\Bzb$ are freely varying parameters.

In the fit, the \CP violation observables are free parameters shared across all fit categories and the \Fi~\cite{LHCb-PAPER-2020-019}, $c_{i}$, \si~\cite{BESIII:2020khq, BESIII:2020hpo} and $\kappa$~\cite{LHCb-PAPER-2015-059} parameters are fixed.

The integrated yields of \mbox{\BsDKst} decays are freely varying parameters in four categories given by the \D decay mode and \KS track type, whilst those of the remaining physics backgrounds are fixed to the results of the global fit. For each of the background components, excluding combinatorial background, the fractional yield in a Dalitz plot bin is fixed.
The effect of \CP violation in interference between the final state paths in \mbox{\BsDKst} and \BsDstKst decays is expected to be small because $r_\Bs \sim 0.02$. Therefore, \Bsb particles are assumed to decay exclusively to \Dz mesons, thus the fraction of these candidates in a Dalitz plot bin is given by \Fi. 
The level of \CP violation in \BDstKst decays is likely at a similar level to the signal, but assigned as zero in the fit due to the very small yield of this decay in the fit range. 
Therefore, the fractional yield of this component in a Dalitz plot bin is \Fmi. A systematic uncertainty is assigned for this assumption as discussed in Sec.~\ref{sec:systematics}. 
For the \BDpipi candidates, the \D meson is assumed to be an equal mixture of \Dz and \Dzb mesons because either pion could be misidentified. 
Therefore, the fraction of these decays in a Dalitz plot bin is $0.5(\Fi + \Fmi)$.
The \BDk background is \CP violating and its distribution over the Dalitz plot is therefore parameterised similarly to Eqs.~(\ref{eq:N_Bz}) and (\ref{eq:N_Bzb}) using values of the \CP violation observables determined from Ref.~\cite{LHCb-PAPER-2021-033}, with $\kappa =1$.
Finally, the Dalitz plot distribution of combinatorial background is unknown, thus the corresponding yield in each bin is a free parameter.

After correcting for small biases (the largest of which is 12\% of the statistical uncertainty) and uncertainty undercoverage (the largest inflation was 3\%) using pseudoexperiments, the \CP violation observables are measured to be \mbox{$x_+ = 0.074 \pm 0.086$}, \mbox{$x_- = -0.215 \pm 0.086$}, \mbox{$y_+ =-0.336 \pm 0.105$} and \mbox{$y_- = -0.012 \pm 0.128$}, with the statistical correlation coefficients displayed in the Appendix.
The left plot in Fig.~\ref{fig:results_plots} displays the 68.3\% and 95.5\% confidence regions for the \CP violation observables determined by scanning the profile likelihood function. 
The opening angle between the lines joining the points \mbox{(\xp, \yp)} and \mbox{(\xm, \ym)} with the origin corresponds to $2\gamma$. To understand the distribution of signal across the Dalitz plot the raw asymmetry, $N_i(\Bzb) - N_{-i}(\Bz)/N_i(\Bzb) + N_{-i}(\Bz)$, is calculated for each effective bin pair. An effective bin labelled $i$, is defined to compare the yield of \Bzb decays in a bin $i$ with the yield of \Bz decays in a bin $-i$. Fig.~\ref{fig:results_plots} displays the asymmetries calculated using the binned yields from the default fit, and for illustrative purposes, those determined in an alternative fit where the signal yield in each region of the Dalitz plot is a free parameter. The good agreement between the yield of signal in each bin determined from the \CP violation observables and those determined from the alternative fit demonstrates that Eqs.~(\ref{eq:N_Bz}) and (\ref{eq:N_Bzb}) are an appropriate model for the data. It is possible to see regions of the Dalitz plot where the asymmetry does deviate from zero. However, \CP violation in this measurement is not yet established with the current precision.  

\begin{figure}[tb]
    \centering
        \includegraphics[width=0.4\textwidth]{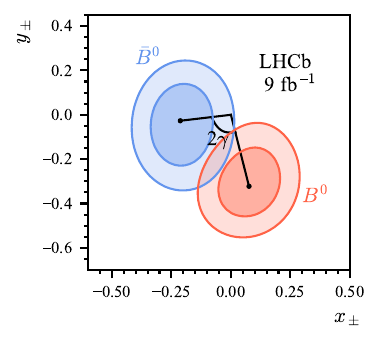}
        \includegraphics[width=0.58\textwidth]{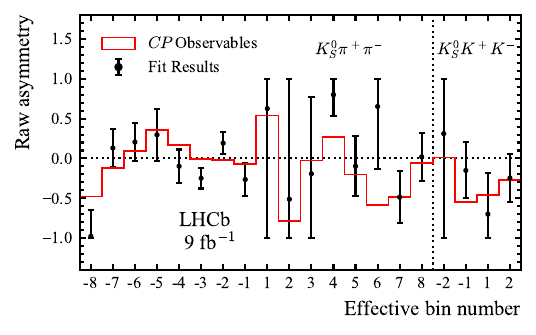}
    \caption{Left: two-dimensional 68.3\% and 95.5\% statistical confidence regions for the measured $(x_\pm, y_\pm)$ values, determined by scanning the profile likelihood function. The orange (blue) contours correspond to the observables related to \Bz (\Bzb) decays. Right: raw asymmetry in each effective bin pair. It is determined using the fitted \CP violation observables (red histogram) and the results of an alternative fit where the signal yield in each Dalitz plot bin is a free parameter (black  points, with statistical uncertainties that are capped to the physical limits where appropriate).}
\label{fig:results_plots}
\end{figure}

\section{Systematic uncertainties}
\label{sec:systematics}
A summary of the systematic uncertainties is displayed in Table~\ref{tab:systematics_ggsz}.
These are primarily evaluated with two methods: the fit to the data is repeated many times using a model with fixed parameters smeared according to their uncertainties and the root-mean-square (RMS) of the \CP violation observable distributions are taken to be the uncertainties, or many pseudodata sets tuned to the data are fitted using a model with an alternative configuration and the biases in the \CP violation observable distributions are taken to be the uncertainty.

Two systematic uncertainties are associated with the \D decay strong-phase inputs.
The effect of their finite precision is determined by generating a set of $c_{i}$ and \si values smeared according to their uncertainties and correlations. 
The corresponding systematic uncertainties are 0.005, 0.004, 0.017 and 0.024 for \xp, \xm, \yp and \ym, respectively.
These are larger for \ypm because the \si values are known less precisely than those of $c_i$, but they remain significantly smaller than the statistical uncertainties.

An uncertainty arises because the effect of $\eta({\mmsq,\mpsq})$ is not accounted for in the \D decay strong-phase parameters.
Alternative $c_{i}$ and \si are calculated using an amplitude model~\cite{BaBar:2018cka} with a flat efficiency profile ($c_i^{\rm{flat}}, \ s_i^{\rm{flat}}$) and an efficiency profile determined using simulated signal decays ($c_i^{\rm{eff}}, \ s_i^{\rm{eff}}$). 
The subsequent systematic uncertainty is evaluated by fitting the data many times using a model with alternative $c_{i}$ and \si coefficients that are generated from a Gaussian with a width equal to the efficiency correction: \mbox{$\delta c_i = c_i^{\rm{flat}} - c_i^{\rm{eff}}$} and \mbox{$\delta s_i = s_i^{\rm{flat}} - s_i^{\rm{eff}}$}.

Selection differences between \mbox{\BDKst} and \BDpi candidates can alter the relative efficiencies in each Dalitz plot bin, introducing a bias on the $\Fi$ parameters appropriate for these decay channels. 
The ratio of the squared $(\pip\pim)_\D$ invariant-mass distribution in simulated \mbox{\BDKst} and \BDpi decays are used to produce an alternative efficiency profile.
This is subsequently applied to an amplitude model~\cite{BaBar:2018cka} to compute different \Fi values. The relative efficiency differences between signal and \BsDKst\ decays are negligible.
It is the dominant systematic uncertainty for the \xpm observables, but is significantly lower than the equivalent uncertainty determined in Ref.~\cite{LHCb-PAPER-2016-006} where the efficiency profile from \mbox{$\Bz\to\Dstar^-\mu^+\nu_\mu$} decays was used.

Various systematic uncertainties related to the fit model are computed.
The dominant contributions are the choice of signal shape and the effect of fixing the combinatorial background slope, signal mean and resolution to the global fit results, and are both evaluated using alternative models.
In the former a different signal distribution is used, and in the latter the slopes in each of the four global fit categories are freely varying parameters that are shared between Dalitz plot bins.
The remaining fit model systematics are those associated with the fixed background ratios, which are evaluated using sets of parameters smeared according to their uncertainties.

In the fit model, \CP violation in partially reconstructed \BDstKst decays is neglected because there are few candidates in the fit range. 
The effect of this assumption is measured using an alternative model where these candidates have the same distribution in phase space as the signal decays.
The amplitude ratio and strong-phase difference for these decays are unknown, but a similar interference as in \mbox{\BDKst} decays is expected since they have a similar amplitude ratio.
Hence, the \CP violation observables determined by the nominal fit are used and the resulting uncertainty is small.

The systematic uncertainty associated with the limited knowledge of the coherence factor, $\kappa$, is determined to be small using an alternative model where its value is displaced by one standard deviation, $\kappa_{\rm{model}} = \kappa - \sigma(\kappa)$~\cite{LHCb-PAPER-2015-059}. 
Larger values of $\kappa$ are not included since its uncertainty is heavily asymmetric, and the lower uncertainty is found to dominate the spread of \CP violation observables.

In the selection, a requirement is placed on the displacement of the \D meson vertex from the \B meson vertex to reduce background from \B decays to the final state particles without the intermediate \D meson, which are referred to as charmless candidates.
Studies of the \D meson invariant-mass sideband determine that the total charmless yield in the sample is $17 \pm 9$.
A systematic uncertainty is assigned using an alternative model where charmless candidates are introduced in the signal region. The yields of these candidates are distributed uniformly over the Dalitz plot and given the small expected yields it is unecessary to account for potential \CP violation in the charmless decays.

Measurements of the Dalitz plot coordinates are affected by the detector momentum resolution and can cause candidates to be assigned to the wrong bin.
To first order, the \Fi values account for this, but the net migration between Dalitz plot bins can differ in \BDpi and \mbox{\BDKst} decays since they exhibit different levels of \CP violation. The expected difference in the \Fi values in \BDpi and \BDKst due to these second order effects is determined using the momentum resolution in simulation, the \CP violation observables of \BDpi~\cite{LHCb-PAPER-2021-033} and those of this analysis and the \D decay model from Ref.~\cite{BaBar:2018cka}. The expected differences are used to generate pseudoexperiments which are then fit with the nominal procedure to assign the systematic uncertainty due to momentum resolution. 

The corrections applied to the \CP violation observables in Sec.~\ref{sec:fit} depend on the physics inputs used in the pseudodata studies.
Therefore, a systematic uncertainty is assigned. 
The values of and correlations between $\gamma$, \rB and \deltaB from Ref.~\cite{LHCb-PAPER-2021-033} are used to generate sets of alternative input \CP violation observables. 
The bias study is repeated many times to create a distribution of corrections, the RMS of which corresponds to the systematic uncertainty.

\begin{table}
\centering
\caption{Systematic uncertainties for the \CP violation observables. Statistical uncertainties are given for reference.}
\label{tab:systematics_ggsz}
\begin{tabular}{lcccc}
\toprule
Source & $\sigma(x_+)$ & $\sigma(x_-)$ & $\sigma(y_+)$ & $\sigma(y_-)$ \\
\midrule
Efficiency correction of $(c_i,s_i)$ & 0.001 & 0.001 & 0.002 & 0.001 \\
$F_i$ inputs & 0.006 & 0.007 & 0.001 & 0.000 \\
Mass Fit & 0.002 & 0.006 & 0.005 & 0.004 \\
$\Bz \to \Dstar\Kstarz$ \CP violation & 0.001 & 0.001 & 0.001 & 0.001 \\
Value of $\kappa$ & 0.000 & 0.001 & 0.003 & 0.002 \\
Charmless background & 0.009 & 0.008 & 0.000 & 0.005 \\
Bin migration & 0.001 & 0.001 & 0.000 & 0.002 \\
Fitter bias & 0.003 & 0.003 & 0.006 & 0.004 \\
\midrule
Total of above systematics & 0.011 & 0.013 & 0.009 & 0.011 \\ \midrule
Strong-phase measurements & 0.005 & 0.004 & 0.017 & 0.024 \\ \midrule
Statistical uncertainty & 0.086 & 0.086 & 0.105 & 0.128 \\
\bottomrule
\end{tabular}
\end{table}

The total systematic uncertainties from all sources excluding those associated with the limited knowledge of the $c_i$ and \si coefficients is determined by summing all the contributions in quadrature. 
They are $0.011$, $0.013$, $0.009$ and $0.011$ for \xp, \xm, \yp and \ym, respectively, and their correlations are given in the Appendix.

\section{Interpretation}
\label{sec:interp}
The \CP violation observables are determined to be
\begin{center}
    \setlength{\tabcolsep}{2pt}
    \renewcommand{\arraystretch}{1.1}
    \begin{tabular}{rcrcrcrcr}
     $x_+$ & $=$ & $0.074$ & $\pm$ & $0.086$ & $\pm$ & $0.005$ & $\pm$ & $0.011$, \\
    $x_-$ & $=$ & $-0.215$ & $\pm$ & $0.086$  & $\pm$ & $0.004$ & $\pm$ & $0.013$, \\
    $y_+$ & $=$ & $-0.336$ & $\pm$ & $0.105$  & $\pm$ & $0.017$ & $\pm$ & $0.009$, \\
    $y_-$ & $=$ & $-0.012$ & $\pm$ & $0.128$ & $\pm$ & $0.024$ & $\pm$ & $0.011$, \\
    \end{tabular}
\end{center}
where the first uncertainty is statistical, the second is the systematic contribution from the \D decay strong-phase inputs and the third is from the experimental systematic uncertainties.
The measured \CP violation observables are used in a maximum likelihood fit to determine the physics parameters $\gamma$, \rB and \deltaB. 
The \CP violation observables are invariant under the transformation $\gamma \to \gamma+180\degrees$ and $\deltaB\to\deltaB+180\degrees$ which leads to two unambiguous solutions for the physics observables.
In the region where $0<\gamma<180\degrees$ is satisfied, the best fit values are
\begin{center}
    \setlength{\tabcolsep}{2pt}
    \renewcommand{\arraystretch}{1.2}
    \begin{tabular}{lcrl}
        $\gamma$ & $=$ & $(49^{+ 22}_{-19})$\degrees, \\
        $r_\Bz$ & $=$ & $0.271^{+ 0.065}_{-0.066}$, \\
        $\delta_\Bz$ & $=$ & $(236^{+19}_{-21})$\degrees,
    \end{tabular}
\end{center}
where the uncertainties are calculated using a frequentist method described in Ref.~\cite{LHCb-PAPER-2021-033}. The corresponding 68.3\% and 95.5\% confidence regions in the $\gamma$ vs. \rB and $\gamma$ vs. \deltaB planes are displayed in Fig.~\ref{fig:twodscans}.

\begin{figure}[t]
    \centering
    \setlength{\tabcolsep}{0pt}
        \includegraphics[width=0.48\textwidth]{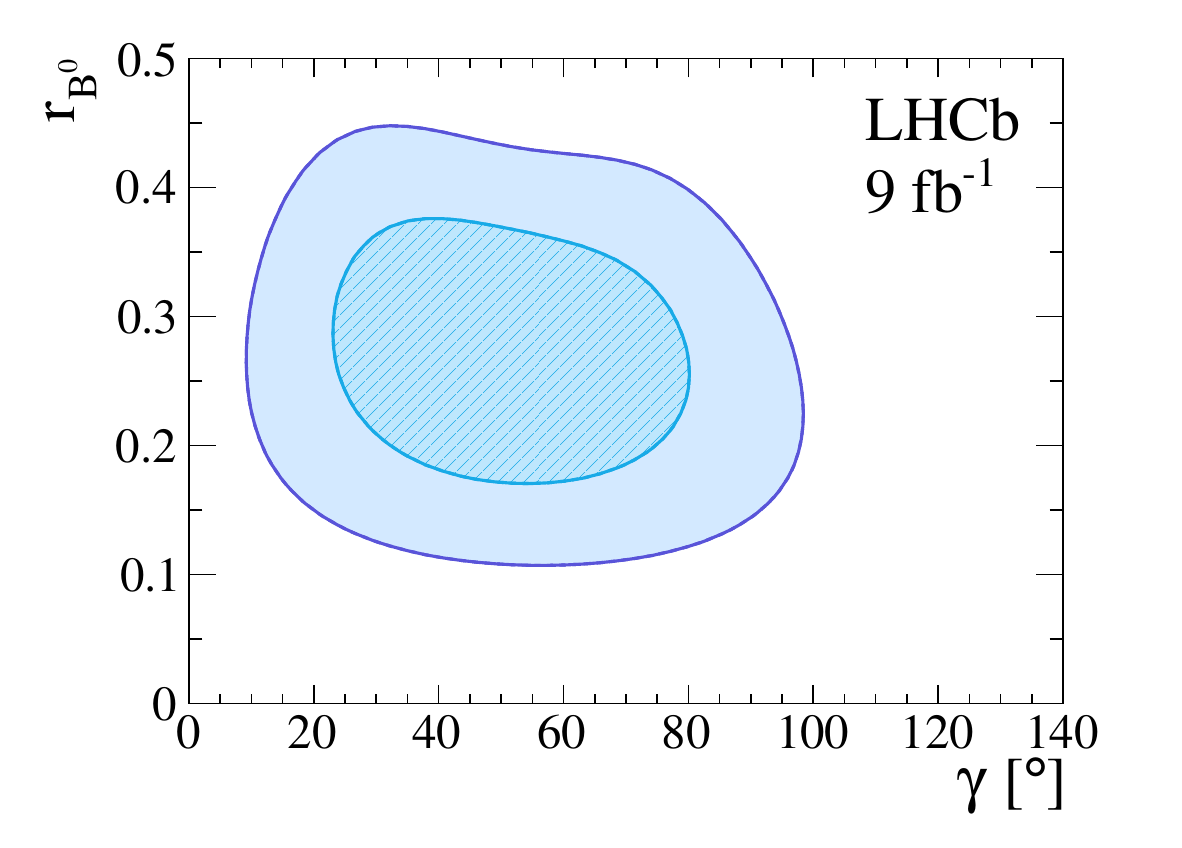}
        \includegraphics[width=0.48\textwidth]{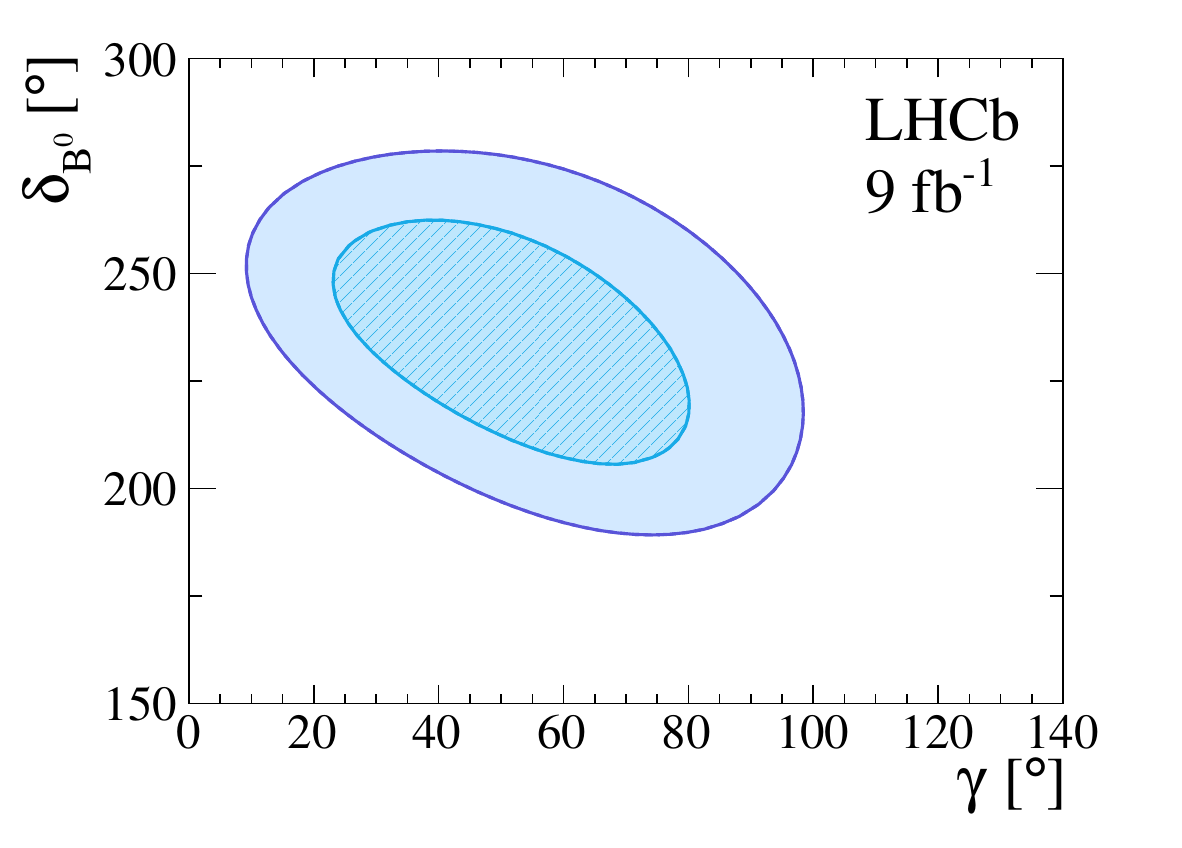}
    \caption{Profile likelihood contours for (left) $\gamma$ versus \rB and (right) $\gamma$ versus \deltaB corresponding to 68.3\% and 95.5\% confidence regions.}
\label{fig:twodscans}
\end{figure}

In the most recent combination of \lhcb results~\cite{LHCb-PAPER-2021-033}, the mean value of $\gamma$ determined using the \mbox{\BDKst} channel was \mbox{$\gamma=(82.0^{+8.1}_{-8.8})\degrees$}, which is higher than the average value using $\B^\pm$ decays, \mbox{$\gamma=(61.7^{+4.4}_{-4.8})\degrees$}. 
The value of $\gamma$ presented here is in good agreement with the current \lhcb average, \mbox{$\gamma = (65.4^{+3.8}_{-4.2})\degrees$}~\cite{LHCb-PAPER-2021-033}, and will reduce the difference between measurements performed using different \bquark-mesons.
Furthermore, it is compatible with the value measured in Ref.~\cite{LHCb-PAPER-2016-006}, \mbox{$\gamma=(71 \pm 20)\degrees$}, although there is not a substantial precision improvement despite using a larger data set. 
This is explained by noting that the uncertainty on $\gamma$ is inversely proportional to the value of \rB, which had a higher central value in Ref.~\cite{LHCb-PAPER-2016-006} than the current measurement.
The value of \rB\ presented in this paper is consistent with previous determinations from \lhcb~\cite{LHCb-PAPER-2019-021}, \babar~\cite{PhysRevD.80.031102} and \belle~\cite{Belle:2012bcb, Belle:2015roy}.
The precision of the \CP violation observables have significantly improved and therefore the results presented here will have a larger weight in future $\gamma$ combinations.

\section{Summary}
\label{sec:conclusion}
Proton-proton collision data corresponding to an integrated luminosity of 9\invfb collected by the \lhcb experiment at centre-of-mass energies of $\sqrt{s}=7, \, 8$ and $13$\tev are used to perform a binned, model-independent \CP violation study of \mbox{\BDKst} decays to measure the CKM angle $\gamma$.
Strong-phase information of \mbox{\Dhh} decays (where $h = \pi,\ K$) from the CLEO~\cite{CLEO:2010iul} and BESIII~\cite{BESIII:2020hlg, BESIII:2020khq, BESIII:2020hpo} experiments is used as inputs.
The measured value is \mbox{$\gamma = (49^{+22}_{-19})\degrees$}, where the uncertainty is statistically dominated and systematic contributions are an order of magnitude
smaller. 
The \CP violation observables measured here are consistent with and supersede those presented in Ref.~\cite{LHCb-PAPER-2016-006}. 


\clearpage
\newpage

\section*{Appendix: Correlation matrices}

\appendix
\label{app:correlations}
Tables \ref{tab:ggsz_correlation_stat} and \ref{tab:ggsz_correlation_syst_nocisi} display the correlation coefficients between the statistical and systematic uncertainties (excluding the strong-phase inputs) on the \CP violation observables, respectively.

A systematic uncertainty is assigned to account for the finite precision on the \D decay strong-phase inputs, $c_i$ and \si~\cite{BESIII:2020khq, BESIII:2020hpo}.
It is given by the RMS of the distributions of \CP violation observables obtained from fitting the data many times using a model with $c_i$ and \si values that are smeared according to their uncertainties and correlations.
This procedure is common between model-independent $\gamma$ measurements.
Therefore, the alternative $c_i$ and \si used for this study are taken from an analysis of \BDh decays at \lhcb~\cite{LHCb-PAPER-2020-019}, which allows correlation coefficients between \CP violation observables of both analysis to be computed. Thus, in combinations the correlation of this systematic uncertainty can be accounted for. 
These are displayed in Table \ref{tab:b2dh_cisi_correlation}.

\begin{table}[h]
    \centering
    \caption{Statistical correlation matrix for the \CP violation observables.}
    \begin{tabular}{l|rrrr} & $x_+$ & $x_-$ & $y_+$ & $y_-$ \\
    \midrule
    $x_+$ & $1.00$ & $0.00$ & $0.18$ & $0.00$ \\
    $x_-$ &  & $1.00$ & $0.00$ & $0.08$ \\
    $y_+$ &  &  & $1.00$ & $0.00$ \\
    $y_-$ &  &  &  & $1.00$ \\
    \end{tabular}
    \label{tab:ggsz_correlation_stat}
\end{table}

\begin{table}[h]
    \centering
    \caption{Correlations between the \CP violation observables for systematic uncertainties excluding the strong-phase inputs.}
    \begin{tabular}{l|rrrr} & $x_+$ & $x_-$& $y_+$ & $y_-$ \\
    \midrule
        $x_+$ & {1.00} & {$-$0.02} & {0.05} & {0.00}  \\
        $x_-$ &  & {1.00} & {0.05} & {0.04}  \\
        $y_+$ &  &  & {1.00} & {$-$0.07}  \\
        $y_-$ &  &  &  & {1.00} \\
    \end{tabular}
    \label{tab:ggsz_correlation_syst_nocisi}
\end{table}

\begin{table}[h]
    \centering
    \caption{Correlations in the \CP violation observables for the strong-phase related systematic uncertainties in \BDKst and \BDh~\cite{LHCb-PAPER-2020-019}.}
    \begin{tabular}{l|rrrrrrrrrr}
        & $x_+^{D\Kstarz}$ & $x_-^{D\Kstarz}$ & $y_+^{D\Kstarz}$ & $y_-^{D\Kstarz}$ & $x_-^{DK}$ & $x_+^{DK}$ & $y_-^{DK}$ & $y_+^{DK}$ & $x_\xi^{D\pi}$ & $y_\xi^{D\pi}$ \\
        \midrule
$x_+^{D\Kstarz}$ & {1.00} & {$-$0.14} & {0.34} & {$-$0.09} & {$-$0.29} & {$-$0.06} & {$-$0.11} & {$-$0.06} & {$-$0.16} & {0.30} \\
$x_-^{D\Kstarz}$ & & {1.00} & {$-$0.04} & {0.17} & {$-$0.31} & {0.48} & {0.22} & {$-$0.49} & {0.04} & {0.12} \\
$y_+^{D\Kstarz}$ & & & {1.00} & {$-$0.04} & {0.35} & {0.03} & {0.12} & {0.27} & {0.29} & {0.32} \\
$y_-^{D\Kstarz}$ & & & & {1.00} & {0.13} & {$-$0.15} & {0.22} & {$-$0.01} & {0.21} & {0.36} \\
$x_-^{DK}$       & & & & & {1.00} & {$-$0.49} & {$-$0.05} & {0.32} & {0.19} & {0.14} \\
$x_+^{DK}$       & & & & & & {1.00} & {0.06} & {0.06} & {0.00} & {$-$0.14} \\
$y_-^{DK}$       & & & & & & & {1.00} & {$-$0.24} & {$-$0.12} & {$-$0.12} \\
$y_+^{DK}$       & & & & & & & & {1.00} & {0.12} & {$-$0.20} \\
$x_\xi^{D\pi}$   & & & & & & & & & {1.00} & {0.64} \\
$y_\xi^{D\pi}$   & & & & & & & & & & {1.00} \\
    \end{tabular} 
    \label{tab:b2dh_cisi_correlation}
\end{table}
\clearpage
\newpage

\clearpage
\newpage
\section*{Acknowledgements}
%
%
\noindent We express our gratitude to our colleagues in the CERN
accelerator departments for the excellent performance of the LHC. We
thank the technical and administrative staff at the LHCb
institutes.
We acknowledge support from CERN and from the national agencies:
CAPES, CNPq, FAPERJ and FINEP (Brazil); 
MOST and NSFC (China); 
CNRS/IN2P3 (France); 
BMBF, DFG and MPG (Germany); 
INFN (Italy); 
NWO (Netherlands); 
MNiSW and NCN (Poland); 
MEN/IFA (Romania); 
MICINN (Spain); 
SNSF and SER (Switzerland); 
NASU (Ukraine); 
STFC (United Kingdom); 
DOE NP and NSF (USA).
We acknowledge the computing resources that are provided by CERN, IN2P3
(France), KIT and DESY (Germany), INFN (Italy), SURF (Netherlands),
PIC (Spain), GridPP (United Kingdom), 
CSCS (Switzerland), IFIN-HH (Romania), CBPF (Brazil),
Polish WLCG  (Poland) and NERSC (USA).
We are indebted to the communities behind the multiple open-source
software packages on which we depend.
Individual groups or members have received support from
ARC and ARDC (Australia);
Minciencias (Colombia);
AvH Foundation (Germany);
EPLANET, Marie Sk\l{}odowska-Curie Actions, ERC and NextGenerationEU (European Union);
A*MIDEX, ANR, IPhU and Labex P2IO, and R\'{e}gion Auvergne-Rh\^{o}ne-Alpes (France);
Key Research Program of Frontier Sciences of CAS, CAS PIFI, CAS CCEPP, 
Fundamental Research Funds for the Central Universities, 
and Sci. \& Tech. Program of Guangzhou (China);
GVA, XuntaGal, GENCAT, Inditex, InTalent and Prog.~Atracci\'on Talento, CM (Spain);
SRC (Sweden);
the Leverhulme Trust, the Royal Society
 and UKRI (United Kingdom).

\bibliographystyle{LHCb}
\bibliography{main,standard,LHCb-PAPER,LHCb-CONF,LHCb-DP,LHCb-TDR}

\ifx\mcitethebibliography\mciteundefinedmacro
\PackageError{LHCb.bst}{mciteplus.sty has not been loaded}
{This bibstyle requires the use of the mciteplus package.}\fi
\providecommand{\href}[2]{#2}
\begin{mcitethebibliography}{10}
\mciteSetBstSublistMode{n}
\mciteSetBstMaxWidthForm{subitem}{\alph{mcitesubitemcount})}
\mciteSetBstSublistLabelBeginEnd{\mcitemaxwidthsubitemform\space}
{\relax}{\relax}

\bibitem{Cabibbo:1963yz}
N.~Cabibbo, \ifthenelse{\boolean{articletitles}}{\emph{{Unitary symmetry and
  leptonic decays}},
  }{}\href{https://doi.org/10.1103/PhysRevLett.10.531}{Phys.\ Rev.\ Lett.\
  \textbf{10} (1963) 531}\relax
\mciteBstWouldAddEndPuncttrue
\mciteSetBstMidEndSepPunct{\mcitedefaultmidpunct}
{\mcitedefaultendpunct}{\mcitedefaultseppunct}\relax
\EndOfBibitem
\bibitem{Kobayashi:1973fv}
M.~Kobayashi and T.~Maskawa,
  \ifthenelse{\boolean{articletitles}}{\emph{{\CP-violation in the
  renormalizable theory of weak interaction}},
  }{}\href{https://doi.org/10.1143/PTP.49.652}{Prog.\ Theor.\ Phys.\
  \textbf{49} (1973) 652}\relax
\mciteBstWouldAddEndPuncttrue
\mciteSetBstMidEndSepPunct{\mcitedefaultmidpunct}
{\mcitedefaultendpunct}{\mcitedefaultseppunct}\relax
\EndOfBibitem
\bibitem{Brod:2013sga}
J.~Brod and J.~Zupan, \ifthenelse{\boolean{articletitles}}{\emph{{The ultimate
  theoretical error on $\gamma$ from $B \to DK$ decays}},
  }{}\href{https://doi.org/10.1007/JHEP01(2014)051}{JHEP \textbf{01} (2014)
  051}, \href{http://arxiv.org/abs/1308.5663}{{\normalfont\ttfamily
  arXiv:1308.5663}}\relax
\mciteBstWouldAddEndPuncttrue
\mciteSetBstMidEndSepPunct{\mcitedefaultmidpunct}
{\mcitedefaultendpunct}{\mcitedefaultseppunct}\relax
\EndOfBibitem
\bibitem{HFLAV21}
Y.~Amhis {\em et~al.}, \ifthenelse{\boolean{articletitles}}{\emph{{Averages of
  $b$-hadron, $c$-hadron, and $\tau$-lepton properties as of 2021}},
  }{}\href{https://doi.org/10.1103/PhysRevD.107.052008}{Phys.\ Rev.\
  \textbf{D107} (2023) 052008},
  \href{http://arxiv.org/abs/2206.07501}{{\normalfont\ttfamily
  arXiv:2206.07501}}, {updated results and plots available at
  \href{https://hflav.web.cern.ch}{{\texttt{https://hflav.web.cern.ch}}}}\relax
\mciteBstWouldAddEndPuncttrue
\mciteSetBstMidEndSepPunct{\mcitedefaultmidpunct}
{\mcitedefaultendpunct}{\mcitedefaultseppunct}\relax
\EndOfBibitem
\bibitem{CKMfitter2015}
CKMfitter group, J.~Charles {\em et~al.},
  \ifthenelse{\boolean{articletitles}}{\emph{{Current status of the standard
  model CKM fit and constraints on \hbox{$\Delta F=2$} new physics}},
  }{}\href{https://doi.org/10.1103/PhysRevD.91.073007}{Phys.\ Rev.\
  \textbf{D91} (2015) 073007},
  \href{http://arxiv.org/abs/1501.05013}{{\normalfont\ttfamily
  arXiv:1501.05013}}, {updated results and plots available at
  \href{http://ckmfitter.in2p3.fr/}{{\texttt{http://ckmfitter.in2p3.fr/}}}}\relax
\mciteBstWouldAddEndPuncttrue
\mciteSetBstMidEndSepPunct{\mcitedefaultmidpunct}
{\mcitedefaultendpunct}{\mcitedefaultseppunct}\relax
\EndOfBibitem
\bibitem{UTfit:2022hsi}
UTfit collaboration, M.~Bona {\em et~al.},
  \ifthenelse{\boolean{articletitles}}{\emph{{New UTfit Analysis of the
  unitarity triangle in the Cabibbo-Kobayashi-Maskawa scheme}},
  }{}\href{https://doi.org/10.1007/s12210-023-01137-5}{Rend.\ Lincei Sci.\
  Fis.\ Nat.\  \textbf{34} (2023) 37},
  \href{http://arxiv.org/abs/2212.03894}{{\normalfont\ttfamily
  arXiv:2212.03894}}\relax
\mciteBstWouldAddEndPuncttrue
\mciteSetBstMidEndSepPunct{\mcitedefaultmidpunct}
{\mcitedefaultendpunct}{\mcitedefaultseppunct}\relax
\EndOfBibitem
\bibitem{Belle:2012bcb}
Belle collaboration, K.~Negishi {\em et~al.},
  \ifthenelse{\boolean{articletitles}}{\emph{{Search for the decay $B^0\to
  DK^{*0}$ followed by $D\to K^-\pi^+$}},
  }{}\href{https://doi.org/10.1103/PhysRevD.86.011101}{Phys.\ Rev.\ D
  \textbf{86} (2012) 011101},
  \href{http://arxiv.org/abs/1205.0422}{{\normalfont\ttfamily
  arXiv:1205.0422}}\relax
\mciteBstWouldAddEndPuncttrue
\mciteSetBstMidEndSepPunct{\mcitedefaultmidpunct}
{\mcitedefaultendpunct}{\mcitedefaultseppunct}\relax
\EndOfBibitem
\bibitem{Belle:2015roy}
Belle collaboration, K.~Negishi {\em et~al.},
  \ifthenelse{\boolean{articletitles}}{\emph{{First model-independent Dalitz
  analysis of $B^0 \to DK^{*0}$, $D\to K_S^0\pi^+\pi^-$ decay}},
  }{}\href{https://doi.org/10.1093/ptep/ptw030}{PTEP \textbf{2016} (2016)
  043C01}, \href{http://arxiv.org/abs/1509.01098}{{\normalfont\ttfamily
  arXiv:1509.01098}}\relax
\mciteBstWouldAddEndPuncttrue
\mciteSetBstMidEndSepPunct{\mcitedefaultmidpunct}
{\mcitedefaultendpunct}{\mcitedefaultseppunct}\relax
\EndOfBibitem
\bibitem{PhysRevD.80.031102}
BABAR Collaboration, B.~Aubert {\em et~al.},
  \ifthenelse{\boolean{articletitles}}{\emph{Search for
  $b\ensuremath{\rightarrow}u$ transitions in
  ${B}^{0}\ensuremath{\rightarrow}{D}^{0}{K}^{*0}$ decays},
  }{}\href{https://doi.org/10.1103/PhysRevD.80.031102}{Phys.\ Rev.\ D
  \textbf{80} (2009) 031102}\relax
\mciteBstWouldAddEndPuncttrue
\mciteSetBstMidEndSepPunct{\mcitedefaultmidpunct}
{\mcitedefaultendpunct}{\mcitedefaultseppunct}\relax
\EndOfBibitem
\bibitem{LHCb-PAPER-2019-021}
LHCb collaboration, R.~Aaij {\em et~al.},
  \ifthenelse{\boolean{articletitles}}{\emph{{Measurement of \CP observables in
  the process \mbox{\decay{\Bz}{\D\Kstarz}} with two- and four-body \D
  decays}}, }{}\href{https://doi.org/10.1007/JHEP08(2019)041}{JHEP \textbf{08}
  (2019) 041}, \href{http://arxiv.org/abs/1906.08297}{{\normalfont\ttfamily
  arXiv:1906.08297}}\relax
\mciteBstWouldAddEndPuncttrue
\mciteSetBstMidEndSepPunct{\mcitedefaultmidpunct}
{\mcitedefaultendpunct}{\mcitedefaultseppunct}\relax
\EndOfBibitem
\bibitem{LHCb-PAPER-2016-006}
LHCb collaboration, R.~Aaij {\em et~al.},
  \ifthenelse{\boolean{articletitles}}{\emph{{Model-independent measurement of
  the {CKM} angle $\gamma$ using $\BDKst$ decays with $\Dpp$ and $\KS\Kp\Km$}},
  }{}\href{https://doi.org/10.1007/JHEP06(2016)131}{JHEP \textbf{06} (2016)
  131}, \href{http://arxiv.org/abs/1604.01525}{{\normalfont\ttfamily
  arXiv:1604.01525}}\relax
\mciteBstWouldAddEndPuncttrue
\mciteSetBstMidEndSepPunct{\mcitedefaultmidpunct}
{\mcitedefaultendpunct}{\mcitedefaultseppunct}\relax
\EndOfBibitem
\bibitem{ggsz1}
A.~Bondar, \ifthenelse{\boolean{articletitles}}{\emph{Proceedings of {BINP}
  special analysis meeting on dalitz analysis, 24-26 sep. 2022, unpublished},
  }{}\relax
\mciteBstWouldAddEndPuncttrue
\mciteSetBstMidEndSepPunct{\mcitedefaultmidpunct}
{\mcitedefaultendpunct}{\mcitedefaultseppunct}\relax
\EndOfBibitem
\bibitem{Bondar:2005ki}
A.~Bondar and A.~Poluektov,
  \ifthenelse{\boolean{articletitles}}{\emph{{Feasibility study of
  model-independent approach to $\phi_3$ measurement using Dalitz plot
  analysis}}, }{}\href{https://doi.org/10.1140/epjc/s2006-02590-x}{Eur.\ Phys.\
  J.\ C \textbf{47} (2006) 347},
  \href{http://arxiv.org/abs/hep-ph/0510246}{{\normalfont\ttfamily
  arXiv:hep-ph/0510246}}\relax
\mciteBstWouldAddEndPuncttrue
\mciteSetBstMidEndSepPunct{\mcitedefaultmidpunct}
{\mcitedefaultendpunct}{\mcitedefaultseppunct}\relax
\EndOfBibitem
\bibitem{Bondar:2008hh}
A.~Bondar and A.~Poluektov, \ifthenelse{\boolean{articletitles}}{\emph{{The use
  of quantum-correlated $D^0$ decays for $\phi_3$ measurement}},
  }{}\href{https://doi.org/10.1140/epjc/s10052-008-0600-z}{Eur.\ Phys.\ J.\ C
  \textbf{55} (2008) 51},
  \href{http://arxiv.org/abs/0801.0840}{{\normalfont\ttfamily
  arXiv:0801.0840}}\relax
\mciteBstWouldAddEndPuncttrue
\mciteSetBstMidEndSepPunct{\mcitedefaultmidpunct}
{\mcitedefaultendpunct}{\mcitedefaultseppunct}\relax
\EndOfBibitem
\bibitem{Giri:2003ty}
A.~Giri, Y.~Grossman, A.~Soffer, and J.~Zupan,
  \ifthenelse{\boolean{articletitles}}{\emph{{Determining $\gamma$ using $B^\pm
  \to \D K^\pm$ with multibody $\D$ decays}},
  }{}\href{https://doi.org/10.1103/PhysRevD.68.054018}{Phys.\ Rev.\ D
  \textbf{68} (2003) 054018},
  \href{http://arxiv.org/abs/hep-ph/0303187}{{\normalfont\ttfamily
  arXiv:hep-ph/0303187}}\relax
\mciteBstWouldAddEndPuncttrue
\mciteSetBstMidEndSepPunct{\mcitedefaultmidpunct}
{\mcitedefaultendpunct}{\mcitedefaultseppunct}\relax
\EndOfBibitem
\bibitem{BESIII:2020khq}
BESIII collaboration, M.~Ablikim {\em et~al.},
  \ifthenelse{\boolean{articletitles}}{\emph{{Model-independent determination
  of the relative strong-phase difference between $D^0$ and
  $\bar{D}^0\rightarrow K^0_{S,L}\pi^+\pi^-$ and its impact on the measurement
  of the CKM angle $\gamma/\phi_3$}},
  }{}\href{https://doi.org/10.1103/PhysRevD.101.112002}{Phys.\ Rev.\ D
  \textbf{101} (2020) 112002},
  \href{http://arxiv.org/abs/2003.00091}{{\normalfont\ttfamily
  arXiv:2003.00091}}\relax
\mciteBstWouldAddEndPuncttrue
\mciteSetBstMidEndSepPunct{\mcitedefaultmidpunct}
{\mcitedefaultendpunct}{\mcitedefaultseppunct}\relax
\EndOfBibitem
\bibitem{BESIII:2020hpo}
BESIII collaboration, M.~Ablikim {\em et~al.},
  \ifthenelse{\boolean{articletitles}}{\emph{{Improved model-independent
  determination of the strong-phase difference between $D^{0}$ and
  $\bar{D}^{0}\to K^{0}_{\mathrm{S,L}}K^{+}K^{-}$ decays}},
  }{}\href{https://doi.org/10.1103/PhysRevD.102.052008}{Phys.\ Rev.\ D
  \textbf{102} (2020) 052008},
  \href{http://arxiv.org/abs/2007.07959}{{\normalfont\ttfamily
  arXiv:2007.07959}}\relax
\mciteBstWouldAddEndPuncttrue
\mciteSetBstMidEndSepPunct{\mcitedefaultmidpunct}
{\mcitedefaultendpunct}{\mcitedefaultseppunct}\relax
\EndOfBibitem
\bibitem{BESIII:2020hlg}
BESIII collaboration, M.~Ablikim {\em et~al.},
  \ifthenelse{\boolean{articletitles}}{\emph{{Determination of Strong-Phase
  Parameters in $D\rightarrow K^0_{S,L}\pi^+\pi^-$}},
  }{}\href{https://doi.org/10.1103/PhysRevLett.124.241802}{Phys.\ Rev.\ Lett.\
  \textbf{124} (2020) 241802},
  \href{http://arxiv.org/abs/2002.12791}{{\normalfont\ttfamily
  arXiv:2002.12791}}\relax
\mciteBstWouldAddEndPuncttrue
\mciteSetBstMidEndSepPunct{\mcitedefaultmidpunct}
{\mcitedefaultendpunct}{\mcitedefaultseppunct}\relax
\EndOfBibitem
\bibitem{CLEO:2010iul}
CLEO collaboration, J.~Libby {\em et~al.},
  \ifthenelse{\boolean{articletitles}}{\emph{{Model-independent determination
  of the strong-phase difference between $D^0$ and $\bar{D}^0 \to K^0_{S,L} h^+
  h^-$ ($h=\pi,K$) and its impact on the measurement of the CKM angle
  $\gamma/\phi_3$}},
  }{}\href{https://doi.org/10.1103/PhysRevD.82.112006}{Phys.\ Rev.\ D
  \textbf{82} (2010) 112006},
  \href{http://arxiv.org/abs/1010.2817}{{\normalfont\ttfamily
  arXiv:1010.2817}}\relax
\mciteBstWouldAddEndPuncttrue
\mciteSetBstMidEndSepPunct{\mcitedefaultmidpunct}
{\mcitedefaultendpunct}{\mcitedefaultseppunct}\relax
\EndOfBibitem
\bibitem{LHCb-PAPER-2015-059}
LHCb collaboration, R.~Aaij {\em et~al.},
  \ifthenelse{\boolean{articletitles}}{\emph{{Constraints on the unitarity
  triangle angle $\gamma$ from Dalitz plot analysis of
  \mbox{\decay{\Bz}{\D\Kp\pim}} decays}},
  }{}\href{https://doi.org/10.1103/PhysRevD.93.112018}{Phys.\ Rev.\
  \textbf{D93} (2016) 112018}, Erratum
  \href{https://doi.org/10.1103/PhysRevD.94.079902}{ibid.\   \textbf{D94}
  (2016) 079902}, \href{http://arxiv.org/abs/1602.03455}{{\normalfont\ttfamily
  arXiv:1602.03455}}\relax
\mciteBstWouldAddEndPuncttrue
\mciteSetBstMidEndSepPunct{\mcitedefaultmidpunct}
{\mcitedefaultendpunct}{\mcitedefaultseppunct}\relax
\EndOfBibitem
\bibitem{LHCb-PAPER-2020-019}
LHCb collaboration, R.~Aaij {\em et~al.},
  \ifthenelse{\boolean{articletitles}}{\emph{{Measurement of the CKM angle
  $\gamma$ in \mbox{$B^{\pm} \to D K^{\pm}$ and $B^{\pm} \to D \pi^{\pm}$}
  decays with $D \to K_{\rm S} h^+h^-$}},
  }{}\href{https://doi.org/10.1007/JHEP02(2021)169}{JHEP \textbf{02} (2021)
  0169}, \href{http://arxiv.org/abs/2010.08483}{{\normalfont\ttfamily
  arXiv:2010.08483}}\relax
\mciteBstWouldAddEndPuncttrue
\mciteSetBstMidEndSepPunct{\mcitedefaultmidpunct}
{\mcitedefaultendpunct}{\mcitedefaultseppunct}\relax
\EndOfBibitem
\bibitem{BaBar:2005rek}
BaBar collaboration, B.~Aubert {\em et~al.},
  \ifthenelse{\boolean{articletitles}}{\emph{{Measurement of $\gamma$ in $B^\mp
  \to D^{(*)} K^\mp$ decays with a Dalitz analysis of $D \to K^0_S \pi^-
  \pi^+$}}, }{}\href{https://doi.org/10.1103/PhysRevLett.95.121802}{Phys.\
  Rev.\ Lett.\  \textbf{95} (2005) 121802},
  \href{http://arxiv.org/abs/hep-ex/0504039}{{\normalfont\ttfamily
  arXiv:hep-ex/0504039}}\relax
\mciteBstWouldAddEndPuncttrue
\mciteSetBstMidEndSepPunct{\mcitedefaultmidpunct}
{\mcitedefaultendpunct}{\mcitedefaultseppunct}\relax
\EndOfBibitem
\bibitem{PDG2022}
Particle Data Group, R.~L. Workman {\em et~al.},
  \ifthenelse{\boolean{articletitles}}{\emph{{\href{http://pdg.lbl.gov/}{Review
  of particle physics}}}, }{}\href{https://doi.org/10.1093/ptep/ptac097}{Prog.\
  Theor.\ Exp.\ Phys.\  \textbf{2022} (2022) 083C01}\relax
\mciteBstWouldAddEndPuncttrue
\mciteSetBstMidEndSepPunct{\mcitedefaultmidpunct}
{\mcitedefaultendpunct}{\mcitedefaultseppunct}\relax
\EndOfBibitem
\bibitem{LHCb-PAPER-2021-033}
LHCb collaboration, R.~Aaij {\em et~al.},
  \ifthenelse{\boolean{articletitles}}{\emph{{Simultaneous determination of
  CKM~angle~$\gamma$ and charm mixing parameters}},
  }{}\href{https://doi.org/10.1007/JHEP12(2021)141}{JHEP \textbf{12} (2021)
  141}, \href{http://arxiv.org/abs/2110.02350}{{\normalfont\ttfamily
  arXiv:2110.02350}}\relax
\mciteBstWouldAddEndPuncttrue
\mciteSetBstMidEndSepPunct{\mcitedefaultmidpunct}
{\mcitedefaultendpunct}{\mcitedefaultseppunct}\relax
\EndOfBibitem
\bibitem{LHCb-DP-2008-001}
LHCb collaboration, A.~A. Alves~Jr.\ {\em et~al.},
  \ifthenelse{\boolean{articletitles}}{\emph{{The \lhcb detector at the LHC}},
  }{}\href{https://doi.org/10.1088/1748-0221/3/08/S08005}{JINST \textbf{3}
  (2008) S08005}\relax
\mciteBstWouldAddEndPuncttrue
\mciteSetBstMidEndSepPunct{\mcitedefaultmidpunct}
{\mcitedefaultendpunct}{\mcitedefaultseppunct}\relax
\EndOfBibitem
\bibitem{LHCb-DP-2014-002}
LHCb collaboration, R.~Aaij {\em et~al.},
  \ifthenelse{\boolean{articletitles}}{\emph{{LHCb detector performance}},
  }{}\href{https://doi.org/10.1142/S0217751X15300227}{Int.\ J.\ Mod.\ Phys.\
  \textbf{A30} (2015) 1530022},
  \href{http://arxiv.org/abs/1412.6352}{{\normalfont\ttfamily
  arXiv:1412.6352}}\relax
\mciteBstWouldAddEndPuncttrue
\mciteSetBstMidEndSepPunct{\mcitedefaultmidpunct}
{\mcitedefaultendpunct}{\mcitedefaultseppunct}\relax
\EndOfBibitem
\bibitem{BBDT}
V.~V. Gligorov and M.~Williams,
  \ifthenelse{\boolean{articletitles}}{\emph{{Efficient, reliable and fast
  high-level triggering using a bonsai boosted decision tree}},
  }{}\href{https://doi.org/10.1088/1748-0221/8/02/P02013}{JINST \textbf{8}
  (2013) P02013}, \href{http://arxiv.org/abs/1210.6861}{{\normalfont\ttfamily
  arXiv:1210.6861}}\relax
\mciteBstWouldAddEndPuncttrue
\mciteSetBstMidEndSepPunct{\mcitedefaultmidpunct}
{\mcitedefaultendpunct}{\mcitedefaultseppunct}\relax
\EndOfBibitem
\bibitem{Sjostrand:2007gs}
T.~Sj\"{o}strand, S.~Mrenna, and P.~Skands,
  \ifthenelse{\boolean{articletitles}}{\emph{{A brief introduction to PYTHIA
  8.1}}, }{}\href{https://doi.org/10.1016/j.cpc.2008.01.036}{Comput.\ Phys.\
  Commun.\  \textbf{178} (2008) 852},
  \href{http://arxiv.org/abs/0710.3820}{{\normalfont\ttfamily
  arXiv:0710.3820}}\relax
\mciteBstWouldAddEndPuncttrue
\mciteSetBstMidEndSepPunct{\mcitedefaultmidpunct}
{\mcitedefaultendpunct}{\mcitedefaultseppunct}\relax
\EndOfBibitem
\bibitem{Sjostrand:2006za}
T.~Sj\"{o}strand, S.~Mrenna, and P.~Skands,
  \ifthenelse{\boolean{articletitles}}{\emph{{PYTHIA 6.4 physics and manual}},
  }{}\href{https://doi.org/10.1088/1126-6708/2006/05/026}{JHEP \textbf{05}
  (2006) 026}, \href{http://arxiv.org/abs/hep-ph/0603175}{{\normalfont\ttfamily
  arXiv:hep-ph/0603175}}\relax
\mciteBstWouldAddEndPuncttrue
\mciteSetBstMidEndSepPunct{\mcitedefaultmidpunct}
{\mcitedefaultendpunct}{\mcitedefaultseppunct}\relax
\EndOfBibitem
\bibitem{LHCb-PROC-2010-056}
I.~Belyaev {\em et~al.}, \ifthenelse{\boolean{articletitles}}{\emph{{Handling
  of the generation of primary events in Gauss, the LHCb simulation
  framework}}, }{}\href{https://doi.org/10.1088/1742-6596/331/3/032047}{J.\
  Phys.\ Conf.\ Ser.\  \textbf{331} (2011) 032047}\relax
\mciteBstWouldAddEndPuncttrue
\mciteSetBstMidEndSepPunct{\mcitedefaultmidpunct}
{\mcitedefaultendpunct}{\mcitedefaultseppunct}\relax
\EndOfBibitem
\bibitem{Lange:2001uf}
D.~J. Lange, \ifthenelse{\boolean{articletitles}}{\emph{{The EvtGen particle
  decay simulation package}},
  }{}\href{https://doi.org/10.1016/S0168-9002(01)00089-4}{Nucl.\ Instrum.\
  Meth.\  \textbf{A462} (2001) 152}\relax
\mciteBstWouldAddEndPuncttrue
\mciteSetBstMidEndSepPunct{\mcitedefaultmidpunct}
{\mcitedefaultendpunct}{\mcitedefaultseppunct}\relax
\EndOfBibitem
\bibitem{davidson2015photos}
N.~Davidson, T.~Przedzinski, and Z.~Was,
  \ifthenelse{\boolean{articletitles}}{\emph{{PHOTOS interface in C++:
  Technical and physics documentation}},
  }{}\href{https://doi.org/https://doi.org/10.1016/j.cpc.2015.09.013}{Comp.\
  Phys.\ Comm.\  \textbf{199} (2016) 86},
  \href{http://arxiv.org/abs/1011.0937}{{\normalfont\ttfamily
  arXiv:1011.0937}}\relax
\mciteBstWouldAddEndPuncttrue
\mciteSetBstMidEndSepPunct{\mcitedefaultmidpunct}
{\mcitedefaultendpunct}{\mcitedefaultseppunct}\relax
\EndOfBibitem
\bibitem{Allison:2006ve}
Geant4 collaboration, J.~Allison {\em et~al.},
  \ifthenelse{\boolean{articletitles}}{\emph{{Geant4 developments and
  applications}}, }{}\href{https://doi.org/10.1109/TNS.2006.869826}{IEEE
  Trans.\ Nucl.\ Sci.\  \textbf{53} (2006) 270}\relax
\mciteBstWouldAddEndPuncttrue
\mciteSetBstMidEndSepPunct{\mcitedefaultmidpunct}
{\mcitedefaultendpunct}{\mcitedefaultseppunct}\relax
\EndOfBibitem
\bibitem{Agostinelli:2002hh}
Geant4 collaboration, S.~Agostinelli {\em et~al.},
  \ifthenelse{\boolean{articletitles}}{\emph{{Geant4: A simulation toolkit}},
  }{}\href{https://doi.org/10.1016/S0168-9002(03)01368-8}{Nucl.\ Instrum.\
  Meth.\  \textbf{A506} (2003) 250}\relax
\mciteBstWouldAddEndPuncttrue
\mciteSetBstMidEndSepPunct{\mcitedefaultmidpunct}
{\mcitedefaultendpunct}{\mcitedefaultseppunct}\relax
\EndOfBibitem
\bibitem{LHCb-PROC-2011-006}
M.~Clemencic {\em et~al.}, \ifthenelse{\boolean{articletitles}}{\emph{{The
  \lhcb simulation application, Gauss: Design, evolution and experience}},
  }{}\href{https://doi.org/10.1088/1742-6596/331/3/032023}{J.\ Phys.\ Conf.\
  Ser.\  \textbf{331} (2011) 032023}\relax
\mciteBstWouldAddEndPuncttrue
\mciteSetBstMidEndSepPunct{\mcitedefaultmidpunct}
{\mcitedefaultendpunct}{\mcitedefaultseppunct}\relax
\EndOfBibitem
\bibitem{Breiman}
L.~Breiman, J.~H. Friedman, R.~A. Olshen, and C.~J. Stone, {\em Classification
  and regression trees}, Wadsworth international group, Belmont, California,
  USA, 1984\relax
\mciteBstWouldAddEndPuncttrue
\mciteSetBstMidEndSepPunct{\mcitedefaultmidpunct}
{\mcitedefaultendpunct}{\mcitedefaultseppunct}\relax
\EndOfBibitem
\bibitem{AdaBoost}
Y.~Freund and R.~E. Schapire, \ifthenelse{\boolean{articletitles}}{\emph{A
  decision-theoretic generalization of on-line learning and an application to
  boosting}, }{}\href{https://doi.org/10.1006/jcss.1997.1504}{J.\ Comput.\
  Syst.\ Sci.\  \textbf{55} (1997) 119}\relax
\mciteBstWouldAddEndPuncttrue
\mciteSetBstMidEndSepPunct{\mcitedefaultmidpunct}
{\mcitedefaultendpunct}{\mcitedefaultseppunct}\relax
\EndOfBibitem
\bibitem{LHCb-PAPER-2021-043}
LHCb collaboration, R.~Aaij {\em et~al.},
  \ifthenelse{\boolean{articletitles}}{\emph{{Observation of the $B^{0}
  \rightarrow \overline{D}^{*0}K^{+}\pi^{-}$ and \mbox{$B_{s}^{0} \rightarrow
  \overline{D}^{*0}K^{-}\pi^{+}$} decays}},
  }{}\href{https://doi.org/10.1103/PhysRevD.105.072005}{Phys.\ Rev.\
  \textbf{D105} (2022) 072005},
  \href{http://arxiv.org/abs/2112.11428}{{\normalfont\ttfamily
  arXiv:2112.11428}}\relax
\mciteBstWouldAddEndPuncttrue
\mciteSetBstMidEndSepPunct{\mcitedefaultmidpunct}
{\mcitedefaultendpunct}{\mcitedefaultseppunct}\relax
\EndOfBibitem
\bibitem{LHCb-PAPER-2012-037}
LHCb collaboration, R.~Aaij {\em et~al.},
  \ifthenelse{\boolean{articletitles}}{\emph{{Measurement of the fragmentation
  fraction ratio $f_s/f_d$ and its dependence on \B meson kinematics}},
  }{}\href{https://doi.org/10.1007/JHEP04(2013)001}{JHEP \textbf{04} (2013)
  001}, \href{http://arxiv.org/abs/1301.5286}{{\normalfont\ttfamily
  arXiv:1301.5286}}\relax
\mciteBstWouldAddEndPuncttrue
\mciteSetBstMidEndSepPunct{\mcitedefaultmidpunct}
{\mcitedefaultendpunct}{\mcitedefaultseppunct}\relax
\EndOfBibitem
\bibitem{LHCb-PAPER-2020-046}
LHCb collaboration, R.~Aaij {\em et~al.},
  \ifthenelse{\boolean{articletitles}}{\emph{{Precise measurement of the
  $f_s/f_d$ ratio of fragmentation fractions and of $B^0_s$ decay branching
  fractions}}, }{}\href{https://doi.org/10.1103/PhysRevD.104.032005}{Phys.\
  Rev.\  \textbf{D104} (2021) 032005},
  \href{http://arxiv.org/abs/2103.06810}{{\normalfont\ttfamily
  arXiv:2103.06810}}\relax
\mciteBstWouldAddEndPuncttrue
\mciteSetBstMidEndSepPunct{\mcitedefaultmidpunct}
{\mcitedefaultendpunct}{\mcitedefaultseppunct}\relax
\EndOfBibitem
\bibitem{BaBar:2018cka}
BaBar, Belle collaborations, I.~Adachi {\em et~al.},
  \ifthenelse{\boolean{articletitles}}{\emph{{Measurement of $\cos{2\beta}$ in
  $B^{0} \to D^{(*)} h^{0}$ with $D \to K_{S}^{0} \pi^{+} \pi^{-}$ decays by a
  combined time-dependent Dalitz plot analysis of BaBar and Belle data}},
  }{}\href{https://doi.org/10.1103/PhysRevD.98.112012}{Phys.\ Rev.\ D
  \textbf{98} (2018) 112012},
  \href{http://arxiv.org/abs/1804.06153}{{\normalfont\ttfamily
  arXiv:1804.06153}}\relax
\mciteBstWouldAddEndPuncttrue
\mciteSetBstMidEndSepPunct{\mcitedefaultmidpunct}
{\mcitedefaultendpunct}{\mcitedefaultseppunct}\relax
\EndOfBibitem
\end{mcitethebibliography}

\clearpage
\newpage
\centerline
{\large\bf LHCb collaboration}
\begin
{flushleft}
\small
R.~Aaij$^{32}$\lhcborcid{0000-0003-0533-1952},
A.S.W.~Abdelmotteleb$^{51}$\lhcborcid{0000-0001-7905-0542},
C.~Abellan~Beteta$^{45}$,
F.~Abudin{\'e}n$^{51}$\lhcborcid{0000-0002-6737-3528},
T.~Ackernley$^{55}$\lhcborcid{0000-0002-5951-3498},
B.~Adeva$^{41}$\lhcborcid{0000-0001-9756-3712},
M.~Adinolfi$^{49}$\lhcborcid{0000-0002-1326-1264},
P.~Adlarson$^{77}$\lhcborcid{0000-0001-6280-3851},
H.~Afsharnia$^{9}$,
C.~Agapopoulou$^{43}$\lhcborcid{0000-0002-2368-0147},
C.A.~Aidala$^{78}$\lhcborcid{0000-0001-9540-4988}
% ...

{\footnotesize \it

% ...
$^{9}$Universit{\'e} Clermont Auvergne, CNRS/IN2P3, LPC, Clermont-Ferrand, France\\
% ...
$^{32}$Nikhef National Institute for Subatomic Physics, Amsterdam, Netherlands\\
% ...
$^{41}$Instituto Galego de F{\'\i}sica de Altas Enerx{\'\i}as (IGFAE), Universidade de Santiago de Compostela, Santiago de Compostela, Spain\\
% ...
$^{43}$European Organization for Nuclear Research (CERN), Geneva, Switzerland\\
% ...
$^{45}$Physik-Institut, Universit{\"a}t Z{\"u}rich, Z{\"u}rich, Switzerland\\
% ...
$^{49}$H.H. Wills Physics Laboratory, University of Bristol, Bristol, United Kingdom\\
% ...
$^{51}$Department of Physics, University of Warwick, Coventry, United Kingdom\\
% ...
$^{54}$School of Physics and Astronomy, University of Glasgow, Glasgow, United Kingdom\\
$^{55}$Oliver Lodge Laboratory, University of Liverpool, Liverpool, United Kingdom\\
% ...
$^{63}$Syracuse University, Syracuse, NY, United States\\
% ...
$^{77}$Department of Physics and Astronomy, Uppsala University, Uppsala, Sweden, associated to $^{54}$\\
$^{78}$University of Michigan, Ann Arbor, MI, United States, associated to $^{63}$\\
\bigskip
% ...
}
\end{flushleft}

\end{document}